\documentclass[fleqn,usenatbib]{mnras}

\usepackage[T1]{fontenc}
\usepackage{ae,aecompl}
\usepackage{verbatim}

\usepackage{graphicx}	
\usepackage{amsmath}	
\usepackage{amssymb}	
\usepackage{float,subfig}
\usepackage{gensymb}
\usepackage{tablefootnote}

\usepackage[figure,figure*]{hypcap}
\usepackage{newtxtext,newtxmath}

\title[Ultraviolet imaging observations of three jellyfish galaxies]{Ultraviolet imaging observations of three jellyfish galaxies: Star formation suppression in the centre and ongoing star formation in stripped tails}

\author[K. George et al.]{
K. George$^{1}$,\thanks{E-mail:koshyastro@gmail.com}
B. M. Poggianti$^2$,
Neven Tomi{\v{c}}i{\'c}$^{2,3,4}$,
J. Postma$^{5}$,
P. C{\^o}t{\'e}$^{6}$,
J. Fritz$^7$, 
\newauthor  
S. K. Ghosh$^{8}$,
M. Gullieuszik$^2$,
J. B. Hutchings$^{6}$,
A. Moretti$^2$, 
A. Omizzolo$^{9,2}$,
\newauthor  
M. Radovich$^2$,
P. Sreekumar$^{10,11}$
A. Subramaniam$^{11}$,
S.N. Tandon$^{11,12}$,
B. Vulcani$^2$
\\
$^{1}$Faculty of Physics, Ludwig-Maximilians-Universit{\"a}t, Scheinerstr. 1, 81679, Munich, Germany\\
$^{2}$INAF-Astronomical Observatory of Padova,  
vicolo dell'Osservatorio 5   
35122 Padova, Italy\\
$^{3}$ Department of Physics and Astronomy, University of Florence, Via G. Sansone 1, 50019 Sesto Fiorentino, Florence, Italy\\
$^{4}$INAF - Arcetri Astrophysical Observatory, Largo E. Fermi 5, 50127 Firenze, Italy\\
$^{5}$University of Calgary, Calgary, Alberta, Canada\\
$^{6}$National Research Council of Canada, Herzberg Astronomy and Astrophysics Research Centre, Victoria, Canada\\
$^{7}$Instituto de Radioastronomia y Astrofisica, UNAM, Campus
  Morelia, A.P. 3-72, C.P. 58089, Mexico\\ 
$^{8}$Tata Institute of Fundamental Research, Mumbai, India\\
$^{9}$Vatican Observatory, Vatican City, Vatican State\\
$^{10}$ISRO HQ, Antariksh Bhavan, Bangalore 560094, India\\
$^{11}$Indian Institute of Astrophysics, Koramangala II Block, Bangalore, India\\
$^{12}$Inter-University Center for Astronomy and Astrophysics, Pune, India\\}

\date{Accepted XXX. Received YYY; in original form ZZZ}

\pubyear{2018}

\begin{document}
\label{firstpage}
\pagerange{\pageref{firstpage}--\pageref{lastpage}}
\maketitle

\begin{abstract}
Spiral galaxies undergo strong ram-pressure effects when they fall into the galaxy cluster potential. As a consequence, their gas is stripped to form extended tails within which star formation can happen, giving them the typical jellyfish appearance. The ultraviolet imaging observations of jellyfish galaxies provide an opportunity to understand ongoing star formation in the stripped tails. We report the ultraviolet observations of the jellyfish galaxies JW39, JO60, JO194 and  compare with observations in optical continuum and $\mathrm{H}{\alpha}$. We detect knots of star formation in the disk and tails of the galaxies and find that their UV and H$\alpha$ flux are  well correlated. The optical emission line ratio maps of these galaxies are used to identify for every region the emission mechanism, due to either star formation, LINER or a mix of the two phenomena.  The star-forming regions in the emission line maps match very well with the regions having significant UV flux. The central regions of two galaxies (JW39, JO194) show a reduction in UV flux which coincides with composite or LINER regions in the emission line maps. The galaxies studied here demonstrate significant star formation in the stripped tails, suppressed star formation in the central regions and  present a possible case of accelerated quenching happening in jellyfish galaxies.
\end{abstract}

\begin{keywords}
galaxies: clusters: intracluster medium, galaxies: star formation
\end{keywords}



\section{Introduction} \label{sec:intro}

Galaxies accrete hot gas from the intergalactic medium which cools and settles in the disk as neutral and molecular hydrogen. Star formation is fueled by this cold gas, which can be sustained by the continuous supply of gas. There are conditions, either external or internal to the galaxy, that can alter the gas supply or content and affect the molecular hydrogen required for star formation. There are a few processes like stellar feedback, AGN feedback, action of the stellar bar/bulge, which happen internally within the galaxy and do not allow the gas to cool and thus sustain continuous star formation. The external processes like major mergers, ram-pressure stripping and starvation remove or cut down the supply of cold gas in galaxies. These processes have different efficiencies and time scales, but the net effect is the gradual shutdown of star formation within the galaxy, known as star formation quenching. The time scales involved in star formation quenching can change with the environment, galaxy stellar mass as well as the nature of different processes that can act alone or in tandem \citep{Man_2018}.\\

Ram pressure stripping is the process by which gas is removed from spiral galaxies when they fall into galaxy clusters due to the action of the hot plasma present in the immediate vicinity of the galaxy \citep{Gunn_1972}. Spiral galaxies in groups and clusters are observed to undergo this process depending on the infalling galaxy mass and velocity with respect to the host halo and the nature of the intracluster medium (ICM) \citep{Hester_2006}. The stripped gas can form tails of ionised gas extending to several kpc. Spiral galaxies undergoing ram pressure stripping can resemble a jellyfish when observed in ultraviolet (UV) and optical wavebands \citep{Owen_2006,Cortese_2007,Owers_2012,Fumagalli_2014, Ebeling_2014, Rawle_2014, Poggianti_2016, Bellhouse_2017, Gullieuszik_2017,Boselli_2018}. 
The presence of ionised gas in the tails of  such "jellyfish galaxies" can be due to \textit{in situ} star formation, while other processes related to the interaction with the surrounding hot ICM (e.g. thermal conduction, mixing etc) can also happen and are not completely ruled out based on recent observations of selected galaxies \citep{Poggianti_2019a}. Massive, hot and young stars (i.e.,\ OBA spectral types) emit the bulk of their radiation at ultraviolet wavelengths, which makes UV imaging observations a direct probe to study ongoing star formation. Hence, the ongoing star formation in the stripped tails of jellyfish galaxies can be best studied from ultraviolet imaging observations. Previous UV observations of jellyfish galaxies were done on galaxies residing in relatively nearby Coma and Virgo clusters \citep{Chung_2009, Smith_2010,Hester_2010,Fumagalli_2011, Boissier_2012, Kenney_2014,Boselli_2018} but there is a dearth of UV observations of jellyfish galaxies belonging to different clusters. It is not clear whether the star formation along the tails and disk of jellyfish galaxies depend on  the host cluster properties. More observations in UV of such galaxies in different clusters are therefore very much needed to understand the details of ongoing star formation in jellyfish galaxies. \\

The goal of this study is to detect regions of ongoing star formation in the stripped tail and the disk of three jellyfish galaxies observed in UV. One edge-on and two face-on jellyfish galaxies are observed as part of GAs Stripping Phenomena in galaxies with MUSE (GASP) survey. The main aim of GASP survey is to investigate the gas removal process in galaxies using the ionized gas observed with the MUSE integral-field spectrograph on VLT \citep{Poggianti_2017a}. The ongoing star formation in the stripped gas can be directly observed in UV and a better understanding of the progression of star formation in jellyfish galaxies (both on tail and disk) is the motivation for observing selected GASP galaxies in UV. The UV also helps to study any lack of star formation in the disk of the galaxy and correlate that with energetic feedback from active galactic nuclei (AGN) (see \citep{George_2019a}). The H$\alpha$ emission from the stripped tails of jellyfish galaxies could be caused by gas ionised by OB stars with life times $\sim$ 10 Myr, whereas the UV emission is coming from stellar photospheres with lifetimes $\sim$ 200 Myr \citep{Kennicutt_2012}. H$\alpha$ and UV emission thus traces different stellar population ages and a comparison of regions in different wavelengths can give clues to the time scales of star formation in jellyfish galaxies.\\

Previous studies of UV observations of galaxies from the GASP survey have been published for two jellyfish galaxies, a nearly face-on and an edge-on, which give a different view of star formation along the stripped tail \citet{George_2018,Poggianti_2019a}. The observations presented here are from the subsequent observations of three more galaxies. These galaxies are selected based on the spectacular nature of strong ram-pressure stripping as evident from optical B-band, H$\alpha$ observations and also on the viewing constrains imposed by satellite. The three galaxies are assigned with Jclass=5 and 4 typical for extreme jellyfish galaxies \citep{Poggianti_2016}. We have now acquired UVIT observations of more GASP jellyfish galaxies with spectacular stripping features, which will be presented in a followup paper (George et al. in prep).\\

We describe below the UV observations of jellyfish galaxies JO60, JW39 and JO194 along with the details of identification of UV knots along the stripped tail and disk. We compare the $\mathrm{H}{\alpha}$ flux in the detected knots with UV. We discuss the observations in section 2, and present the results in section 3. We summarize the key findings from this study in section 4. Throughout this paper we adopt a Chabrier initial mass function, concordance $\Lambda$ CDM cosmology with $H_{0} = 70\,\mathrm{km\,s^{-1}\,Mpc^{-1}}$, $\Omega_{\rm{M}} = 0.3$, $\Omega_{\Lambda} = 0.7$.

\section{Observations, Data \& Analysis} \label{sec:style}
 
 




\begin{table*}
\centering
\label{galaxy details}
\tabcolsep=0.3cm
\begin{tabular}{cccccccccc} 
\hline 
\hline
ID &  RA &  Dec. & Cluster &  $\sigma_{cl}$ & logM$\star$/M$\odot$ &  z & Jclass & BCG$_{sep}$  & BCG$_{sep}$\\
   & (J2000) & (J2000) &       & (km/s) & & & & (Mpc) &  (R200)  \\
\hline
JO60 & 14:53:51.57 & $+$18:39:06.4  & A1991  &  570 & 10.40 & 0.0622 & 5 &  0.633  & 0.48 \\
JW39 &  13:04:07.71 & $+$19:12:38.5 & A1668  &  654 & 11.21 & 0.0663 & 5 &  0.442  & 0.33 \\
JO194 & 23:57:00.68 & $-$34:40:50.1 & A4059  &  744 & 11.18 & 0.0420 & 4 &  0.286  & 0.17 \\
\hline
\end{tabular}
\caption{\label{t7} Details of three galaxies used in this work. (1) GASP ID taken from \citet{Poggianti_2016}; (2) and (3) equatorial coordinates of the galaxy centre; (4) host galaxy cluster name; (5) velocity dispersion of the host galaxy cluster from \citet{Gullieuszik_2020}; (6) logarithm of the galaxy stellar mass (in solar masses from \citet{Vulcani_2018}); (7) galaxy redshift;  (8) Jellyfish Class from \citet{Poggianti_2016};  (9) \& (10) separation between galaxy and BCG in Mpc and R200}  
\end{table*}

The three galaxies presented in this paper have a spiral morphology and belong to three different galaxy clusters. JO60 is seen almost edge-on and JW39 and JO194 are seen face-on. All the three galaxies are undergoing extreme stripping following the first infall to galaxy cluster according to their position in the phase-space diagram showing the projected infalling velocity normalized for the cluster velocity dispersion as a function of the projected distance from the cluster center \citep{Jaffe_2018}. Details of these galaxies and the host galaxy clusters are given in Table 1 and a description on the host cluster properties are given in Lourenco et al 2022 (submitted). Lourenco et al. (2022) have carried out an in-depth analysis of several indicators (both X-ray and optical) of the dynamical status of clusters. Here we report some of their results for the clusters hosting the three jellyfish galaxies studied here, but we refer to Lourenco et al. (2022) for details. The host clusters of all three jellyfish galaxies show a bright cluster galaxy (BCG) that matches the X-ray center, which is considered a sign of relaxation. A1991 (host of JO60) is a relaxed cluster, with a concentrated and single peaked X-ray emission. A1668 and A4059, instead, have a regular X-ray emission but show small X-ray substructures. In this sense, A1668 and A4059 are similar to the clusters hosting JO201 (A85) and JW100 (A2626), both with small X-ray substructures. Based on the analysis of all WINGS/OMEGAWINGS optical spectroscopic redshifts \citep{Moretti_2017}, all three jellyfishes galaxies analyzed in this paper belong to substructures (Biviano, Private communication), as also JW100 (\citep{Poggianti_2019a}) and JO201 (\citep{Bellhouse_2017}). The projected distance of the three jellyfish galaxies from the BCG are given in Table 1. The  luminosity distance for JO60 is $\sim$ 261 Mpc and the angular scale of 1$\arcsec$ on the sky corresponds to 1.13 kpc at the galaxy cluster rest frame. The  luminosity distance for JW39 is $\sim$ 285 Mpc and the angular scale of 1$\arcsec$ on the sky corresponds to 1.22 kpc at the galaxy cluster rest frame. The  luminosity distance for JO194 is $\sim$ 213 Mpc and the angular scale of 1$\arcsec$ on the sky corresponds to 0.941 kpc at the galaxy cluster rest frame.\\



The galaxy cluster fields hosting JO60, JW39, JO194 were observed at optical wavelengths (B and V passbands) as part of the WINGS and OmegaWINGS surveys \citep{Fasano_2006,Gullieuszik_2015,Moretti_2014,Moretti_2017}. The three jellyfish galaxies were observed with MUSE on the VLT under the programme GASP using the spatially resolved integral field unit spectrograph MUSE \citep{Poggianti_2017a}. We use the MUSE $\mathrm{H}{\beta}$ (4861.33 {\AA}), [OIII] (5006.84 {\AA}), [NII] (6583.45 {\AA}), $\mathrm{H}{\alpha}$ (6562.82 {\AA}) and [SII] (6716.44 {\AA}, 6730.81 {\AA}) emission line flux maps of these galaxies in this study.\\

The UV observations are from the ultra-violet imaging telescope (UVIT) onboard the Indian multi wavelength astronomy satellite AstroSat \citep{Agrawal_2006}. The UVIT consists of twin telescopes, a FUV (130-180nm) telescope and a NUV (200-300nm),VIS (320-550nm) telescope which operate with a dichroic beam splitter. The telescopes are of 38cm diameter and generate circular images over a 28$'$ diameter field simultaneously in all three channels \citep{Kumar_2012}. There are options for a set of narrow and broad band filters, out of which we used the NUV N242W and FUV F148W/F154W  filters for NUV and FUV imaging observations for JO60. UVIT NUV channel stopped working since March, 2018 and we have only FUV imaging observations for JW39 and JO194 \citep{Ghosh_2021}. Table 2 gives details on the UVIT observations of the three jellyfish galaxies.\footnote{PI: Koshy George, Proposal ID: G07${\_}$002 JO60, G08$\_$002 JW39, A05$\_$108 JO194}.\\

The UVIT observations have an angular resolution of $\sim$ 1\farcs2  for the NUV  and  $\sim$ 1\farcs4  for the FUV channels \citep{Annapurni_2016,Tandon_2017a}. The NUV and FUV images are corrected for distortion \citep{Girish_2017}, flat field and satellite drift using the software CCDLAB  \citep{Postma_2017}. The images from multiple orbits are coadded to create the master image. The astrometric calibration is performed using the {\tt astrometry.net} package  where solutions are performed using USNO-B catalog \citep{Lang_2010}. The photometric calibration is done using the zero point values generated for photometric calibration stars as described in  \citet{Tandon_2017b} and updated in \citet{Tandon_2020}. The UV magnitudes are in AB system.\\

The $\mathrm{H}{\alpha}$ emission line flux
map of the three galaxies obtained from MUSE observations, which we mention as $\mathrm{H}{\alpha}$ image for convenience, is used in this study to compare with UV observations. The 1$\arcmin$ $\times$ 1$\arcmin$ field of view requires at least two MUSE pointings to cover the disk and the tail of the galaxies studied here. We note that H$\alpha$ imaging from VLT/MUSE has a plate scale of 0.2 arcsec pixel$^{-1}$ with angular resolution $\sim$ 1$\arcsec$ whereas the UVIT plate scale is 0.4 arcsec pixel$^{-1}$ with angular resolution of $\sim$ 1.2-1.4$\arcsec$ at the position of the galaxy.\\

\begin{table}
\centering
\label{galaxy details}
\tabcolsep=0.05cm
\begin{tabular}{cccccc} 
\hline 
\hline
Galaxy & Channel & Filter & $\lambda_{mean}$({\AA})  &  $\delta$$\lambda$({\AA})  & Int.time(s)  \\
\hline
JO60 &  &        &    &        &  \\
     & FUV  & F148W       &  1481  &   500     &  12900\\
     & NUV  & N242W     &  2418  &   785     &  12516 \\
JW39 &  &        &    &        &  \\
      & FUV  & F154W       &  1541  &   380     &  17989\\
JO194 & & & & & \\
      & FUV  & F148W       &  1481  &   500  & 37504 \\
\hline
\end{tabular}
\caption{\label{t7} Log of UVIT observations}
\end{table}

\section{Results}\label{sec:Results}

\subsection{UV and H$\alpha$ imaging}

The young stars in the stripped tail of jellyfish galaxies can have different ages depending on the time since the start of the stripping and subsequent gas condensation to stars. The UV and H$\alpha$ imaging of jellyfish galaxies with flux coming from different stellar spectral types can then show different morphology. In addition, the H$\alpha$ flux can have a contribution from other ionisation mechanisms like shocks, AGN, ionisation from older stellar generations etc. Below we compare the UV and H$\alpha$ images of the three jellyfish galaxies in our sample. We note that UV traces 100-200 Myr old star formation in FUV and NUV passband imaging used here. We present here a detailed analysis of the UV imaging data for JO60, JW39 and JO194.

\subsubsection{JO60}

\begin{figure*}
\includegraphics[width=1\textwidth]{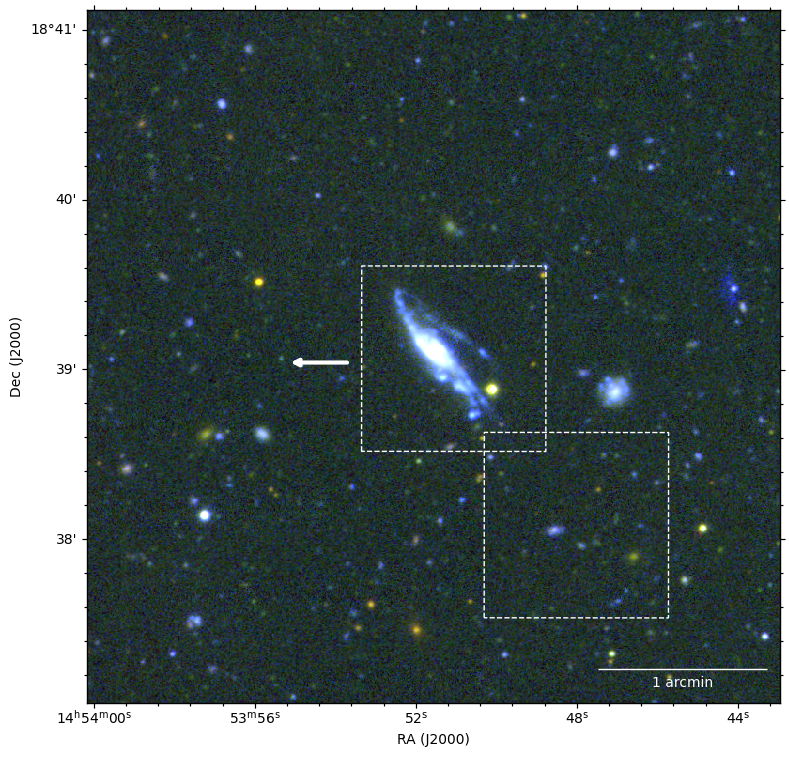}
\caption{The NUV and optical colour composite image of the region around JO60. The NUV image is from observations with UVIT and the optical imaging from the OmegaWINGS survey {\citep{Gullieuszik_2015}}. The image is made by combining NUV (coloured blue) and optical $B,V$ filter band pass images (coloured red and green). 1$\arcmin$ $\times$ 1$\arcmin$ field of view of VLT/MUSE pointings are marked with a white dashed line box. The direction towards the bright cluster galaxy is shown with a white arrow. The image measures 4$\arcmin$ $\times$ 4$\arcmin$ ($\sim$ 271 kpc $\times$ 271 kpc).}\label{figure:JO60comp}
\end{figure*}

\begin{figure*}
\centering
\subfloat[]{\includegraphics[width=0.5\textwidth]{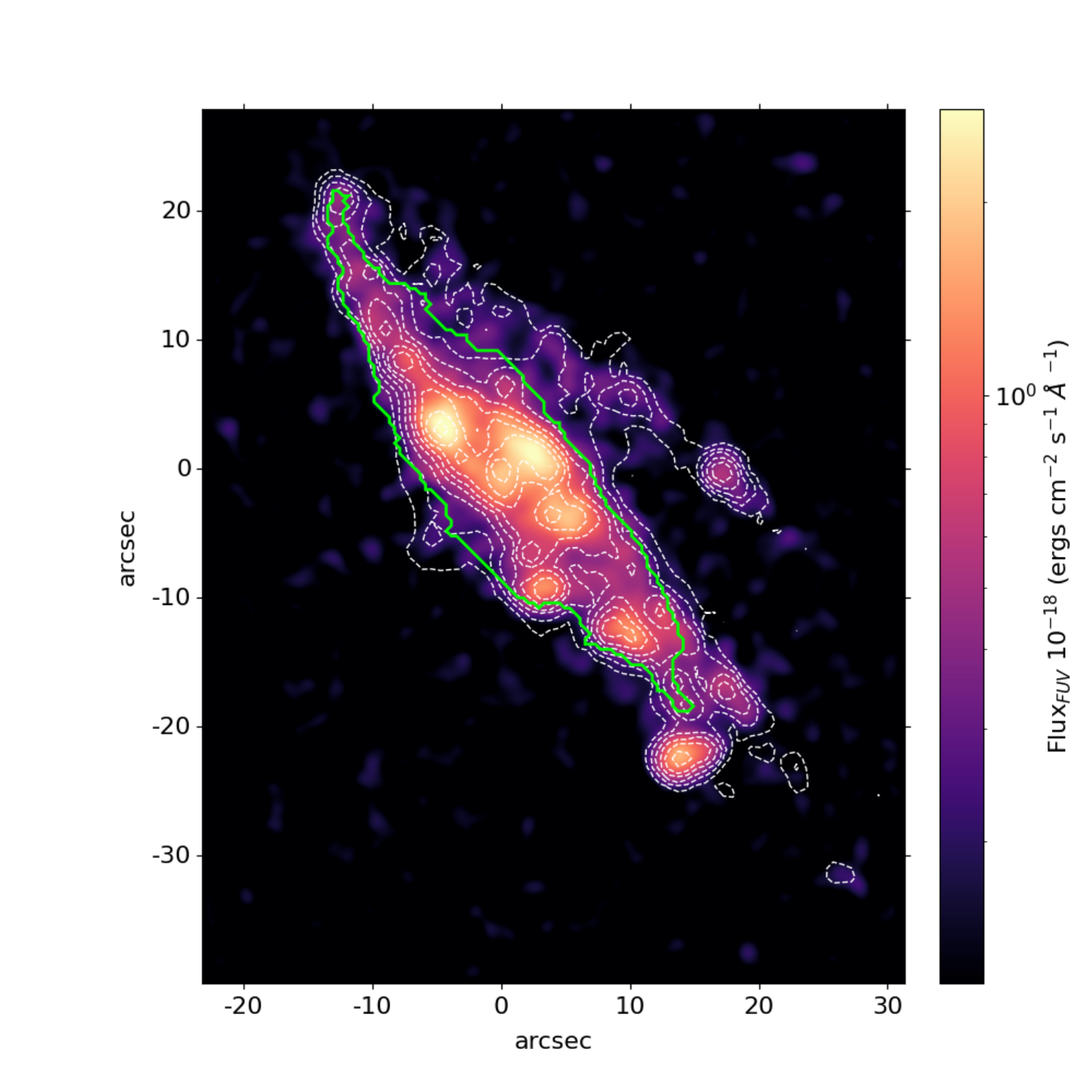}}
\subfloat[]{\includegraphics[width=0.5\textwidth]{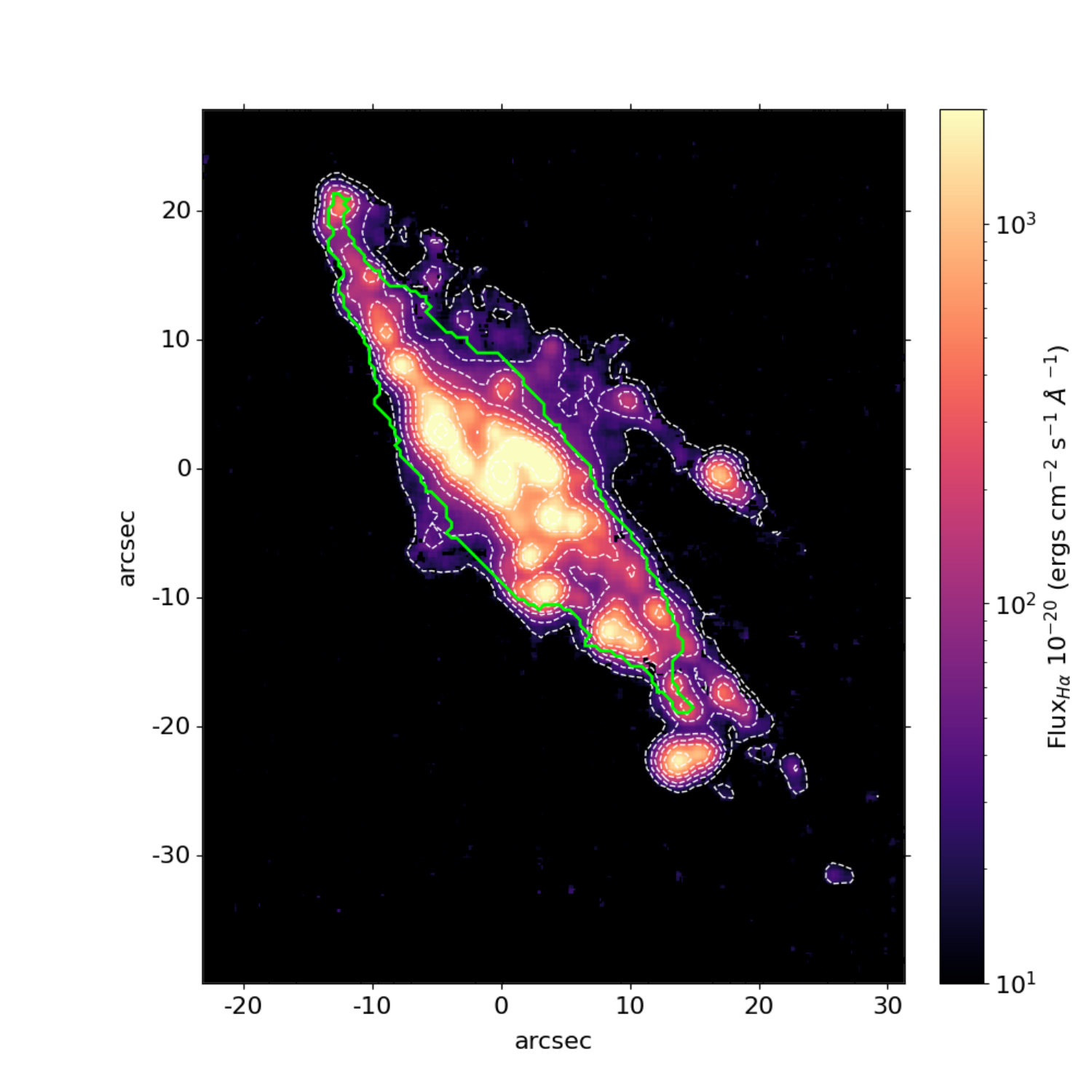}}
\caption{The $\mathrm{H}{\alpha}$ flux contours in white colour is overlaid over the FUV (a) and $\mathrm{H}{\alpha}$ image of JO60 (b). The green line define the galaxy main body as described in \citet{Gullieuszik_2020}.
}\label{figure:JO60fuvhalpha}
\end{figure*}

\begin{figure}
\centering
\includegraphics[width=0.5\textwidth]{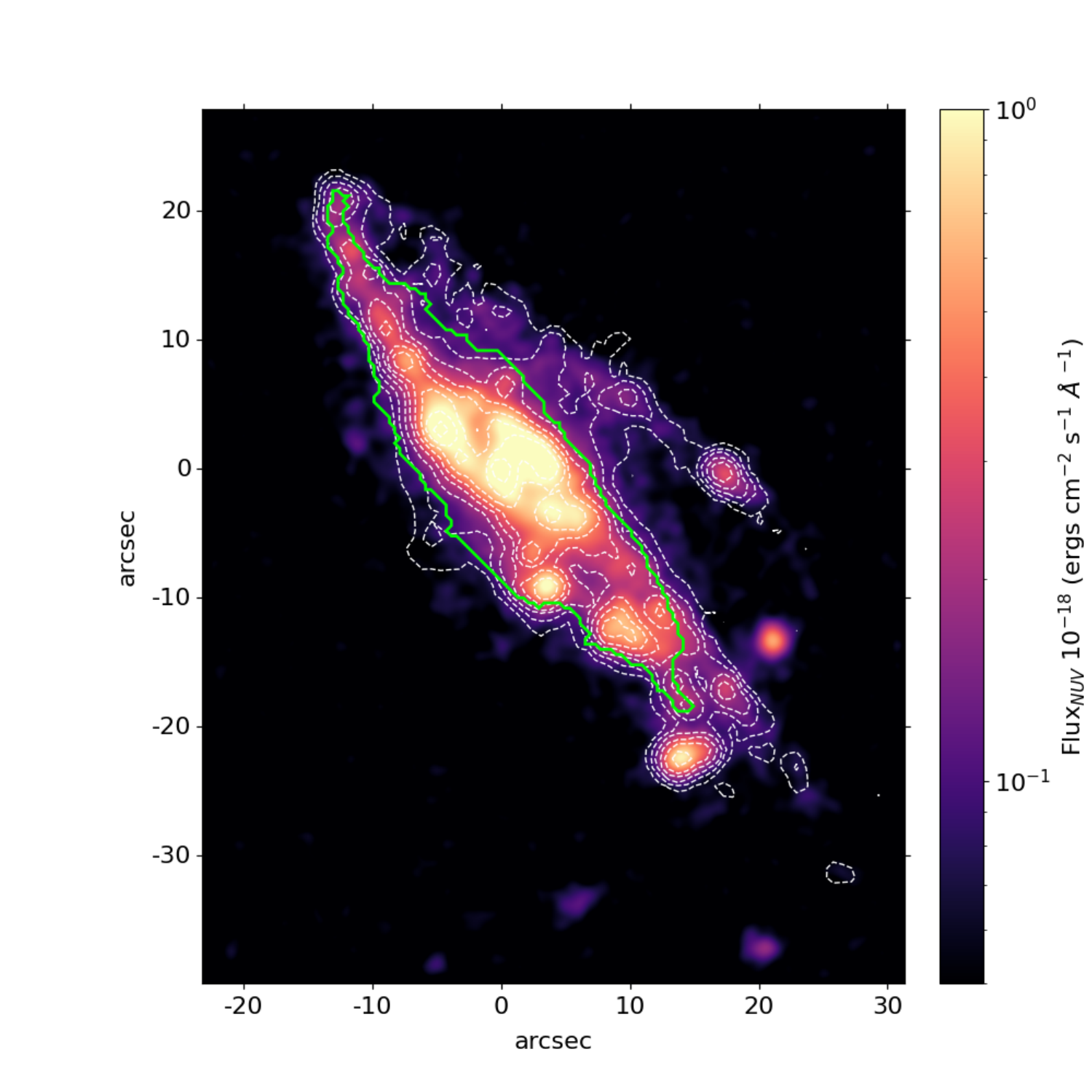}
\caption{The $\mathrm{H}{\alpha}$ flux contours in white colour is overlaid over the NUV image of JO60. The green line define the galaxy main body.}\label{figure:JO60nuv}
\end{figure}

The UVIT observations of Abell 1991 cover a diameter of $\sim$ 1.9 Mpc at the galaxy cluster rest frame. The UV and optical colour composite image of the region centred on JO60 is shown in Figure \ref{figure:JO60comp}. The galaxy is seen almost edge-on with an unwinding pattern of tail at two sides, very similar to the scenario proposed in \citet{Bellhouse_2021} but not very clear due to the edge-on view of the galaxy.  We note a pattern of knots that is visible towards the south. It is not clear whether these knots are part of the system as we lack MUSE coverage for this region. We note that the UV emission along the tails is more diffuse in nature with a lower number of knots comparing to the UV image of the other jellyfish galaxy JO201 studied under GASP (See Fig 2 of \citet{George_2018} where the UV emission is seen more in the form of knots along with diffuse emission close to the disk of the galaxy). We point out that JO60 is seen nearly edge-on and the projection effects should also be taken into account while looking for UV features like knots. Perhaps the knots are present in the orthogonal direction to the line of sight and hence they are not visible when viewed edge-on. However there are few UV knots that could be star forming regions confined to the southern part of the tail of JO60. There is a bright star just on the west of the southern side of the galaxy which is visible in optical as well as NUV (but not in FUV) imaging data. The galaxy seen at the west is not having redshift information to make any association with JO60. The flux calibrated FUV, NUV and H$\alpha$ imaging data of JO60 are shown in Figure \ref{figure:JO60fuvhalpha} and \ref{figure:JO60nuv}. We overlay the contour that defines the galaxy main body (in green) and the H$\alpha$ contours in dashed white colour. The galaxy main body is determined based on galaxy boundary definition of isophotes from the MUSE spectral continuum map as described in section 3.1 of \citet{Gullieuszik_2020}.
The prominent features that are present in the disk and tail are detected in all three images. There are more UV bright knots in the lower tail compared to the tail in the upper part of the galaxy.\\

\subsubsection{JW39}

\begin{figure*}
\includegraphics[width=1\textwidth]{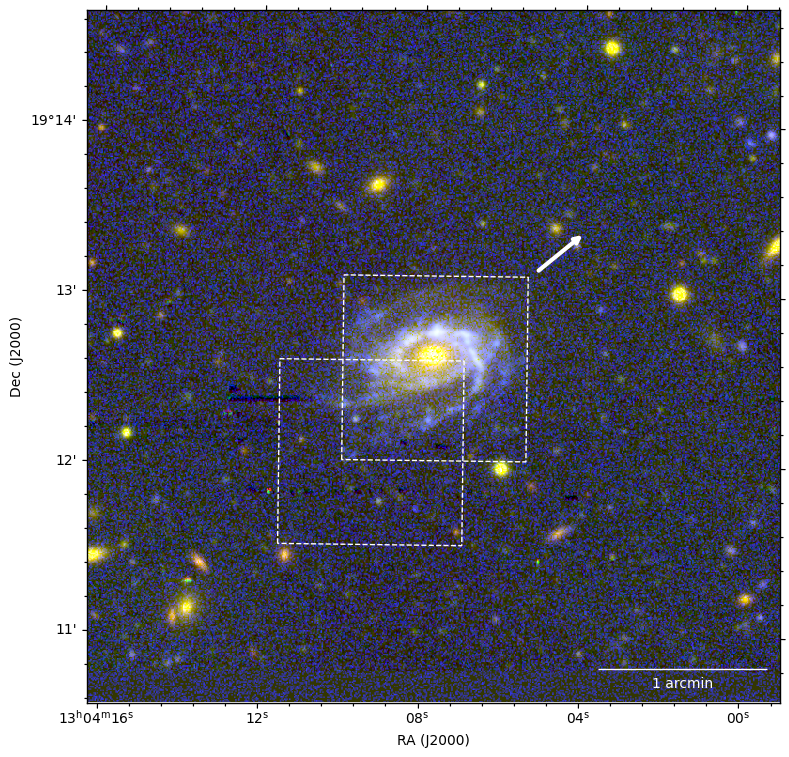}
\caption{The FUV and optical colour composite image of the region around JW39. This colour composite image was made by combining the FUV image from UVIT with optical B,V imaging from the WINGS survey {\citep{Fasano_2006}}. 1$\arcmin$ $\times$ 1$\arcmin$ field of view of VLT/MUSE pointings are marked with a white dashed line box. The direction towards the bright cluster galaxy is shown with a white arrow. The image measures 4$\arcmin$ $\times$4$ \arcmin$ ($\sim$ 293 kpc $\times$ 293kpc).}\label{figure:JW39comp}
\end{figure*}

\begin{figure*}
\centering
\subfloat[]{\includegraphics[width=0.52\textwidth]{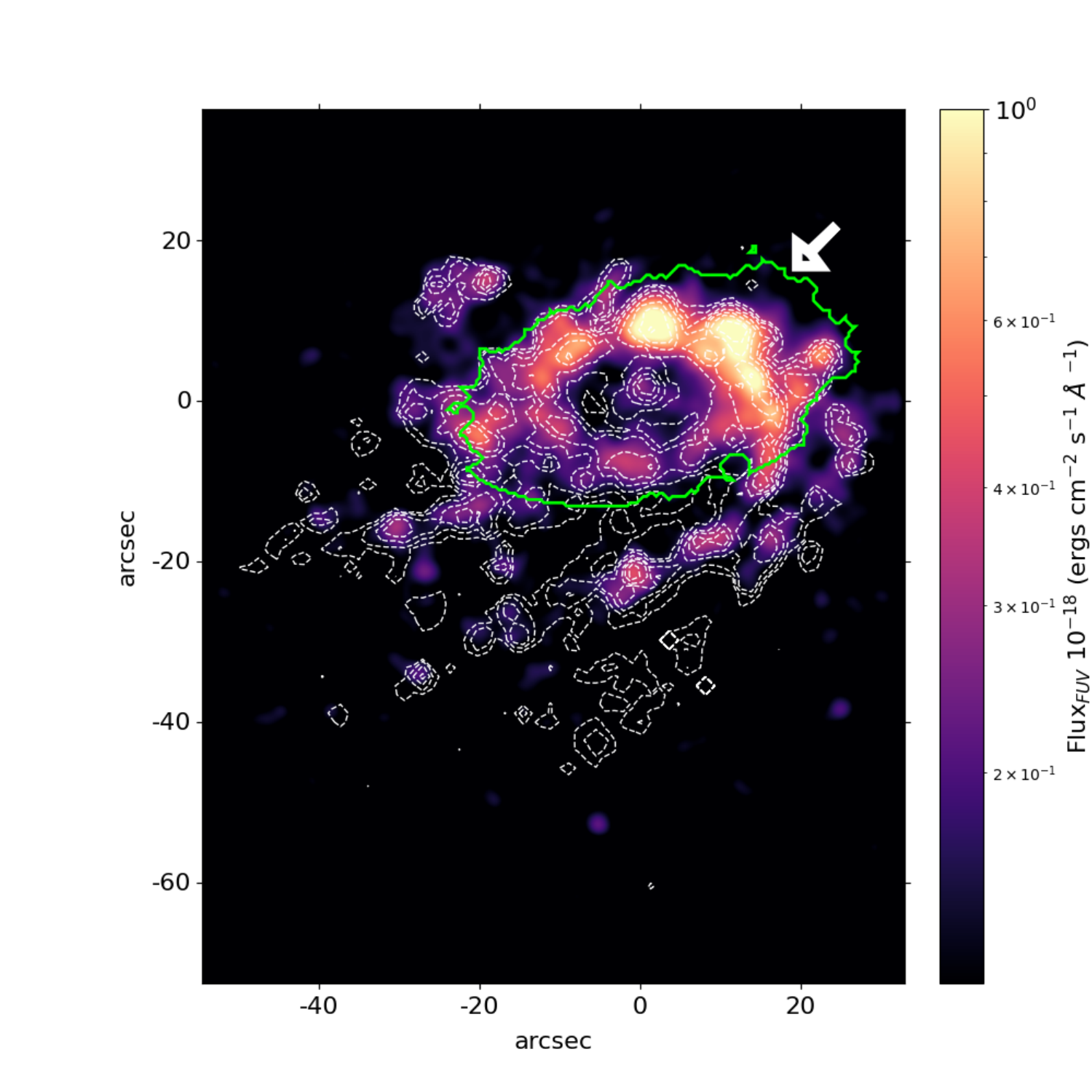}}
\subfloat[]{\includegraphics[width=0.52\textwidth]{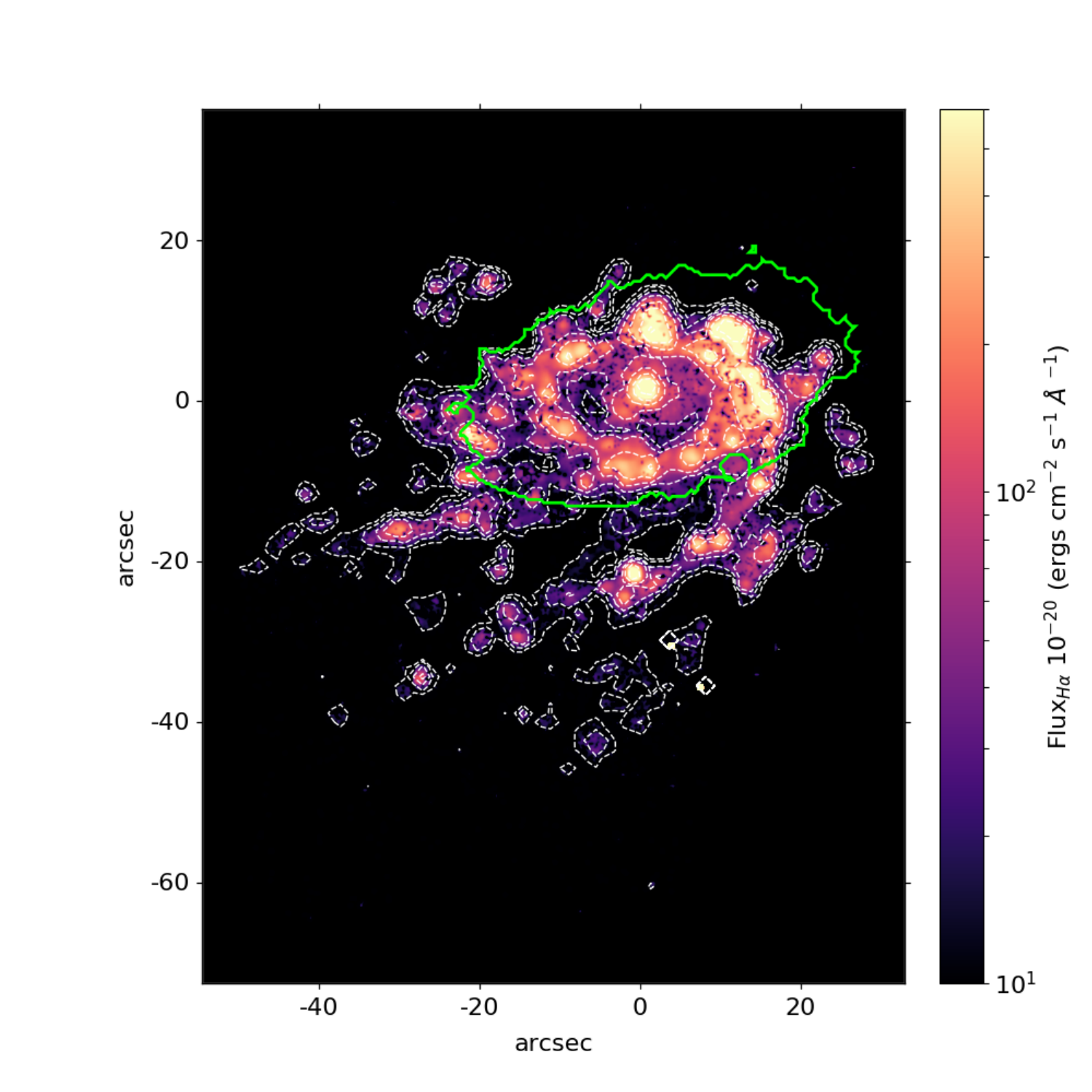}}
\caption{The $\mathrm{H}{\alpha}$ flux contours in white colour is overlaid over the FUV image of JW39 (a) and the $\mathrm{H}{\alpha}$ image of JW39 (b). The green line define the galaxy main body as described in \citet{Gullieuszik_2020}. The white colour arrow points to the enhanced UV emission region and therefore possible interaction site of galaxy disk with the ICM.}\label{figure:JW39fuvhalpha}
\end{figure*}

The UVIT observations of the Abell 1668 galaxy cluster correspond to a diameter of $\sim$ 2 Mpc at the cluster rest frame. The UV and optical colour composite image of the region centred on JW39 is shown in Figure \ref{figure:JW39comp}. The UV emission along the tails of the galaxy is prominent. The flux calibrated FUV and H$\alpha$ imaging data of JW39 are shown in Figure \ref{figure:JW39fuvhalpha}. The central region of the galaxy is dominated in the optical by the presence of a bulge.

The stripping process forms tails that appear as an unwinding of the arms of the galaxy as reported in \citet{Bellhouse_2021}. There are regions on the UV image, particularly at the lower side of the stripped tail with low levels of UV flux contained in H$\alpha$ contour. H$\alpha$ and UV otherwise match very well on the disk of the galaxy. We note that there is a cavity around the central region source seen strikingly in UV and partly also in H$\alpha$. The galaxy shows enhanced UV and H$\alpha$ emission towards the upper side of the disk (marked with a white arrow in Figure \ref{figure:JW39fuvhalpha}a) at the region that most probably is impacting on the ICM during the galaxy infall. The uppermost side of the disk is however already devoid of both UV and H$\alpha$ emission, thus it has been quenched already at least $\sim$ 100-200 Myr ago.

\subsubsection{JO194}

\begin{figure*}
\includegraphics[width=1\textwidth]{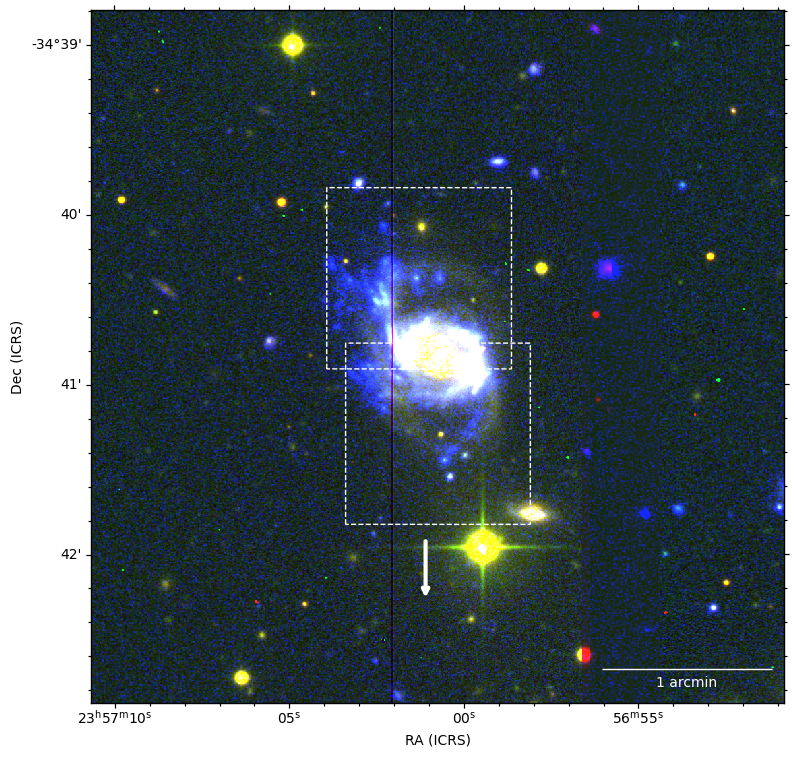}
\caption{The FUV and optical colour composite image of the region around JO194. This colour composite image was made by combining the FUV image from UVIT with optical B,V imaging from the OmegaWINGS survey \citep{Gullieuszik_2015}. 1$\arcmin$ $\times$ 1$\arcmin$ field of view of VLT/MUSE pointings are marked with a white dashed line box. The direction towards the bright cluster galaxy is shown with a white arrow. The image measures 4$\arcmin$ $\times$ 4$\arcmin$ ($\sim$ 226 kpc $\times$ 226kpc). The vertical stripe is due to a gap in optical B and V imaging.}\label{figure:JO194comp}
\end{figure*}

\begin{figure*}
\centering
\subfloat[]{\includegraphics[width=0.52\textwidth]{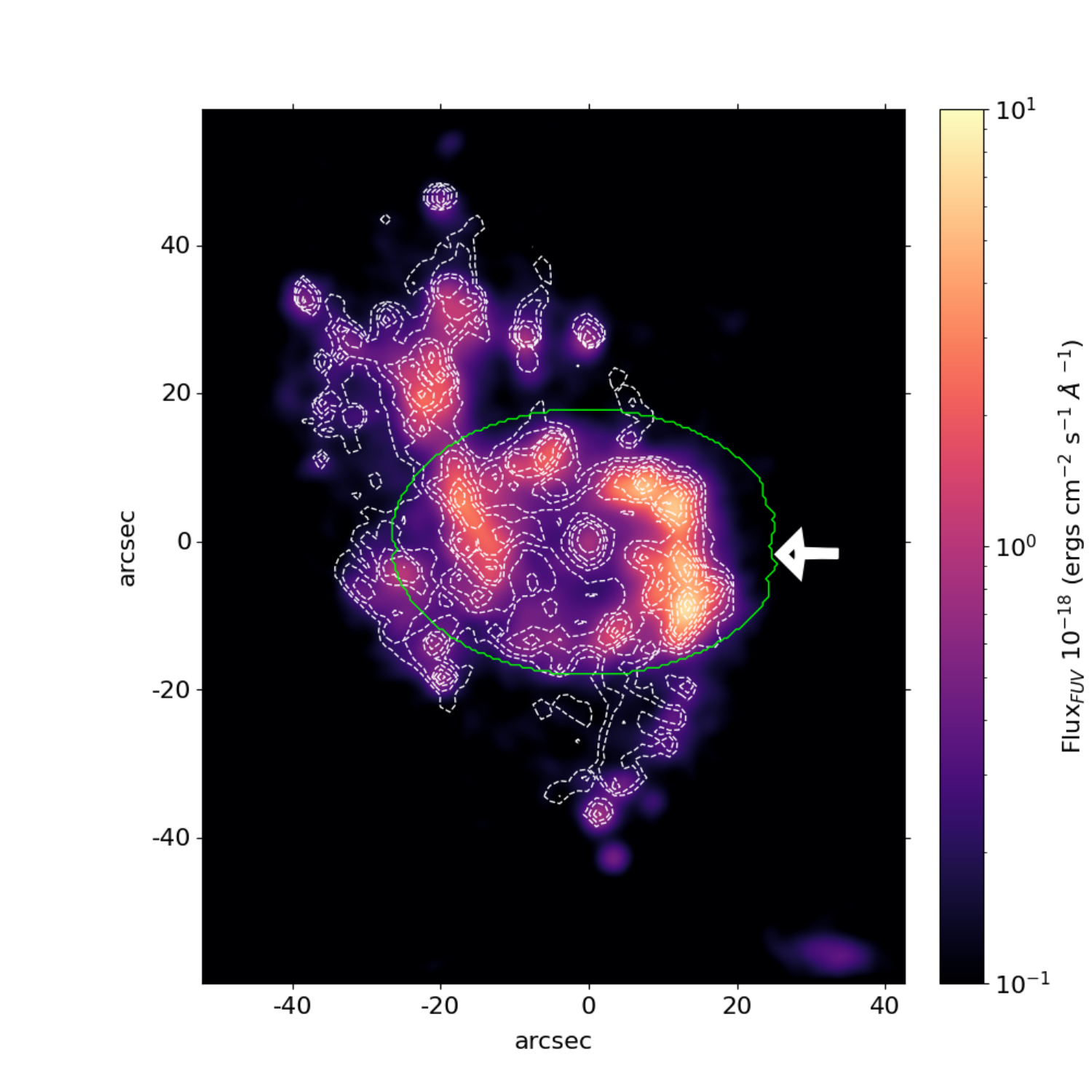}}
\subfloat[]{\includegraphics[width=0.52\textwidth]{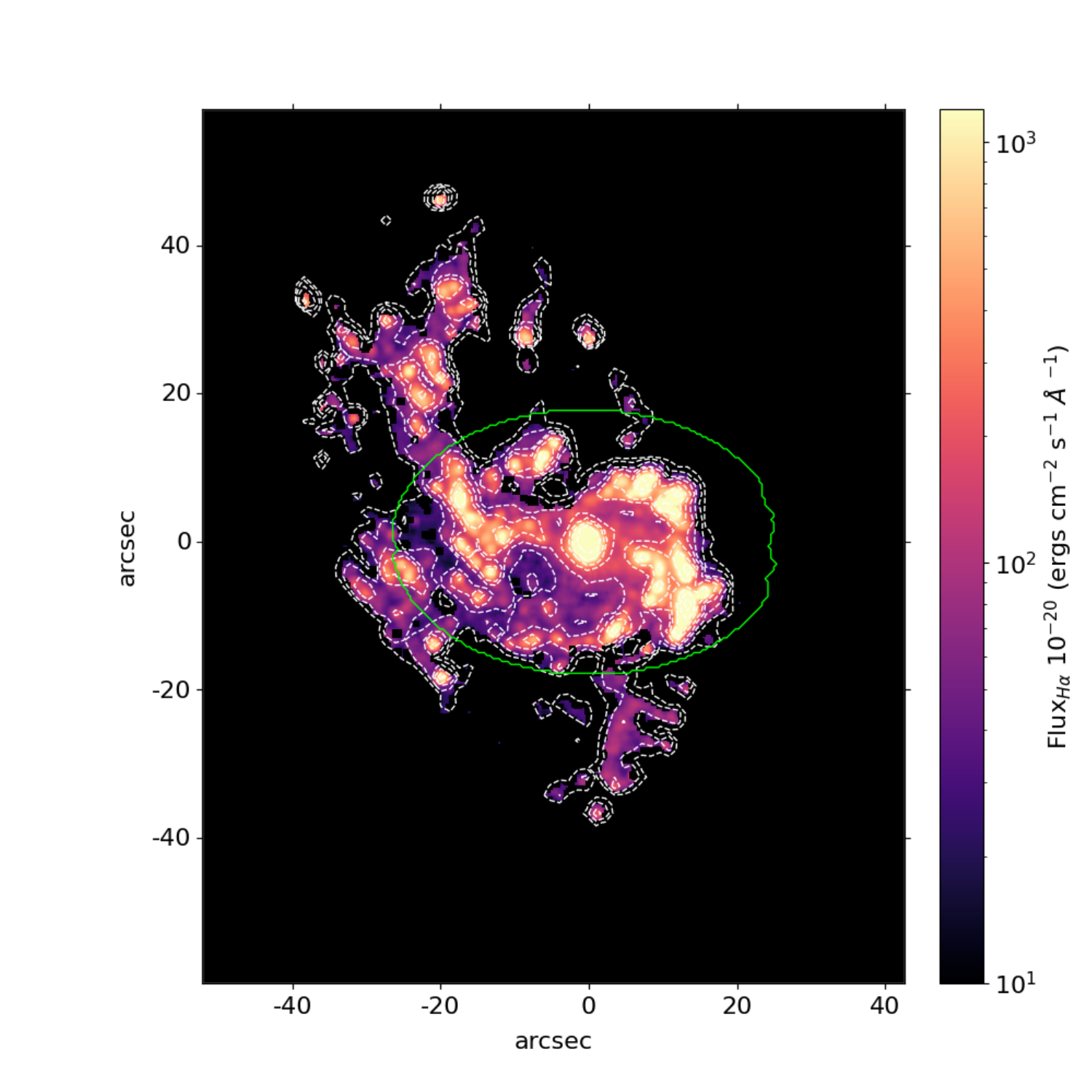}}
\caption{The $\mathrm{H}{\alpha}$ flux contours in white colour is overlaid over the FUV image of JO194 (a) and the $\mathrm{H}{\alpha}$ image of JO194 (b). The green line define the galaxy main body as described in \citet{Gullieuszik_2020}. The white colour arrow points to the enhanced UV emission region and therefore possible interaction site of galaxy disk with the ICM.}\label{figure:JO194fuvhalpha}
\end{figure*}

The UVIT observations of the Abell 4059 galaxy cluster correspond to a diameter of $\sim$ 1.58 Mpc at the galaxy cluster rest frame. The UV and optical colour composite image of the region centred on JO194 is shown in Figure \ref{figure:JO194comp}. The galaxy is seen face on with the stripped tails appearing as an unwinding of the spiral arms \citep{Bellhouse_2021}. The UV emission along the tails is showing knots of star formation along with a diffuse component. There are regions along the arms in the colour composite image in Figure \ref{figure:JO194comp} with no much UV emission but having detectable emission in B,V images. The unwinding arm in the lower side of galaxy is seen as having the optical (B,V) emission slightly displaced from the UV. There is a redder bar situated within the disk seen in optical images. The flux calibrated FUV and H$\alpha$ imaging data of JO194 is shown in Figure \ref{figure:JO194fuvhalpha}. There are regions in the tails, particularly at the lower tail of the galaxy, that are quite bright in H$\alpha$ but not in UV. The spatial offset between UV and H$\alpha$ emission is consistent with the star formation occurring in spiral arms that are progressively opening, as reported for JO194 using spectrophometric modeling in \citet{Bellhouse_2021}. The galaxy shows enhanced UV and H$\alpha$ emission on east side of the disk (marked with a white arrow in Figure \ref{figure:JO194fuvhalpha}a) at the region that most probably interacts first with the ICM during the galaxy infall. This region also shows an offset between the galaxy main body (defined from optical continuum emission) and UV emission suggesting an ongoing quenching propagating along the disk due to ram-pressure stripping over the last 100-200 Myr.

\subsection{Segmentation maps from UV and H$\alpha$ imaging}

\begin{figure*}
\centering
\subfloat[]{\includegraphics[width=0.35\textwidth]{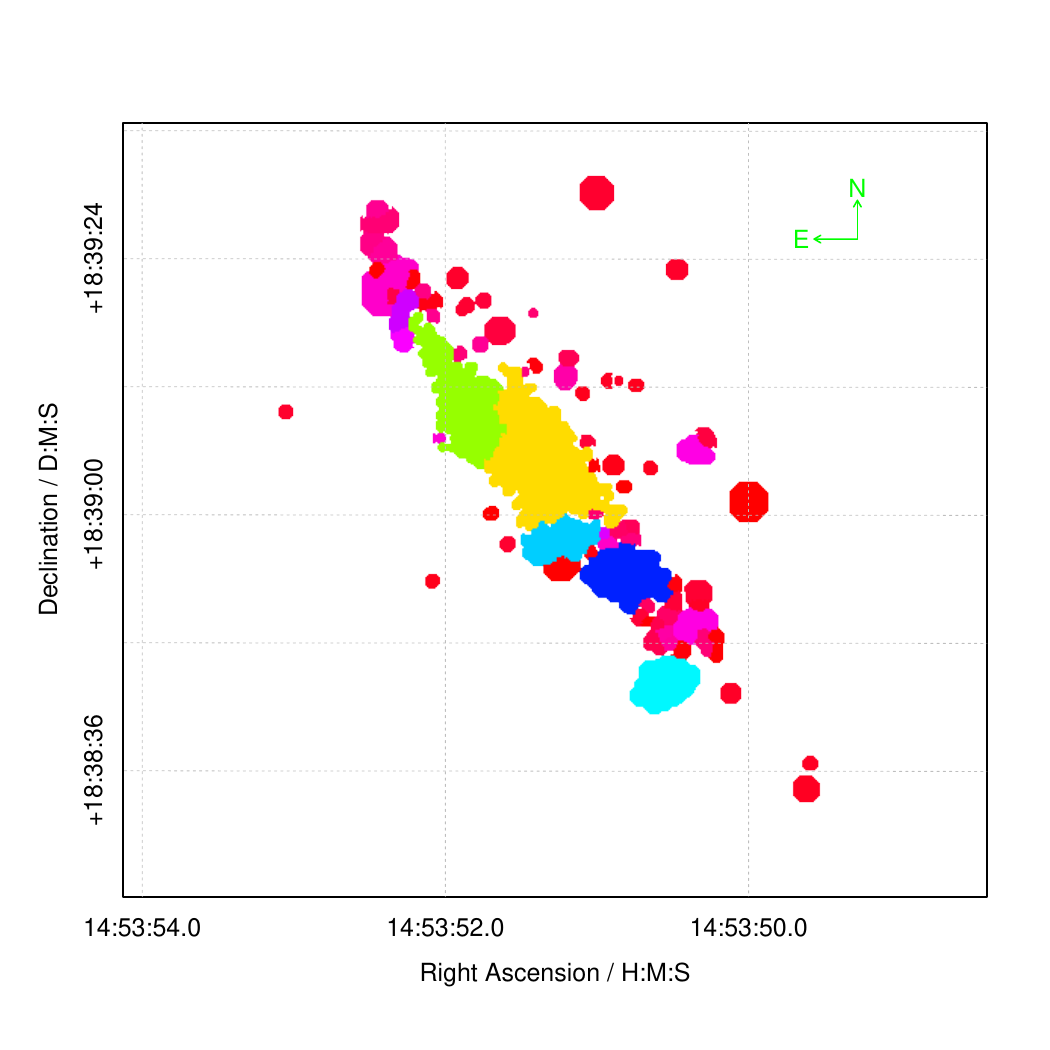}}
\subfloat[]{\includegraphics[width=0.35\textwidth]{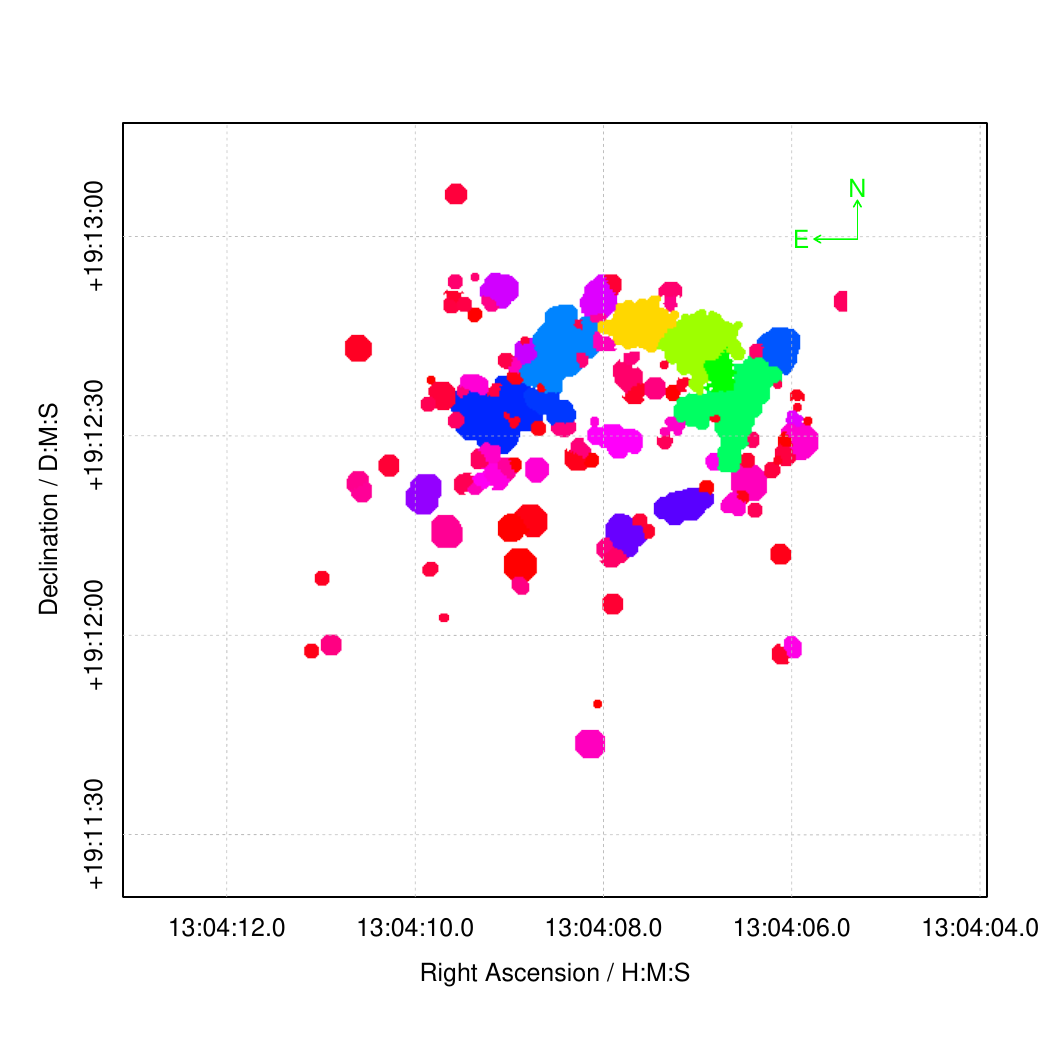}}
\subfloat[]{\includegraphics[width=0.35\textwidth]{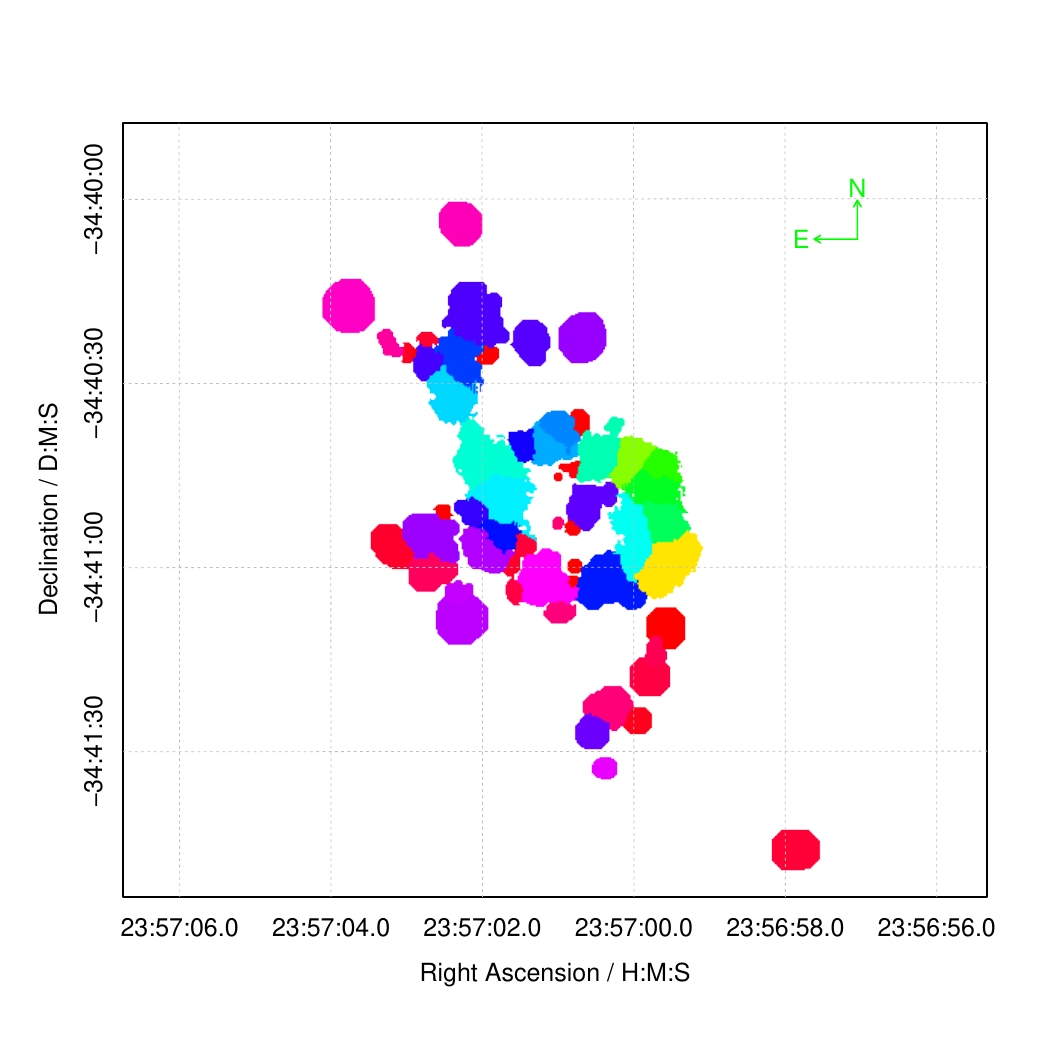}}
\caption{The segmentation map for galaxies JO60, JW39 and JO194 created from ProFound run over FUV image. The colour scheme used to map the segments goes from red for faint segments, to blue and to green for the brightest segments.}\label{figure:segmap}
\end{figure*}



 We used ProFound \citep{Robotham_2018} to identify UV segments from the FUV images of the three galaxies. ProFound is a source finding package that generates dilated segmentation map of pixels belonging to an individual source and performs photometry that fully measures the associated flux from the region. This is a different technique comparing to the traditional approach of using circular or elliptical aperture based photometry. The isophote of the source with irregular shapes called segments are dilated iteratively and converge to measure the background subtracted total flux. The iterative dilation ensures de-blending where segments are not allowed to grow into each other. For the segments thus identified we obtain the central RA, Dec, total flux and  the total number of pixels within which the flux is measured. We first run ProFound on the FUV image of three galaxies creating a segmentation map of the galaxy and identifying various segments on the disk and stripped tails of the galaxy. The positional information of segments from the segmentation map created from FUV image is then used to extract the H$\alpha$ flux in the same region from MUSE data. We note that due to slightly coarser spatial resolution of UVIT images, we decided to use UVIT based segmentation map to measure the total UV and H$\alpha$ fluxes, rather than resampling the Halpha image to the UVIT one. This make sure that the morphological information of the segments are retained and same region position and area is used on the UV and H$\alpha$ images of the galaxy to measure the flux. The segmentation maps created for the three galaxies are shown in Fig \ref{figure:segmap}. The colour scheme used to map the segments  change from red for faint segments, to blue and to green for the brightest segments. The bonafide segments with redshifts belonging to three galaxies are used in further analysis. The segments that are background or foreground objects with different redshifts are isolated based on the redshift information from the MUSE data cube and not used in our analysis. We identify the segments falling within the galaxy main body (shown in green line in Fig \ref{figure:JO60fuvhalpha}, Fig \ref{figure:JW39fuvhalpha}, and Fig \ref{figure:JO194fuvhalpha}) as contributing from the galaxy disk and the rest from the tails.
 \\
 
The ProFound run on the JO60 FUV image detects 77 segments and by careful analysis overlaying the segments on the images, we could find 67 segments in common between FUV and H$\alpha$.  There are 34 segments that are identified from the disk of the galaxy and the remaining 33 belong to the stripped tail. The ProFound run on the JW39 FUV image identifies 115 segments and we could find 87 segments in common between FUV and H$\alpha$. There are 42 segments that are identified from the disk of the galaxy and the remaining 45 belong to the stripped tail of the galaxy. The ProFound run on the JO194 FUV image identifies 59 segments out of which we find 54 segments in common between FUV and H$\alpha$.  There are 32 segments that are identified from the disk of the galaxy and the remaining 22 belong to the stripped tail. Table 3 gives details on the detected segments in common between UV and H$\alpha$ images for the three jellyfish galaxies.

\begin{table}
\centering
\label{galaxy details}
\begin{tabular}{cccc} 
\hline 
\hline
Galaxy &  Disk & Tail & Total  \\
\hline
JO60  &   34    & 33   & 67  \\
JW39  &    42    &  45   &  87  \\
JO194 &    32    &  22   &  54  \\
\hline
\end{tabular}
\caption{\label{t7} Number of segments used in the study detected from UV images of three galaxies.}
\end{table}

\subsection{Comparing UV and H$\alpha$ flux}

\begin{figure}
\centering
\includegraphics[width=0.52\textwidth]{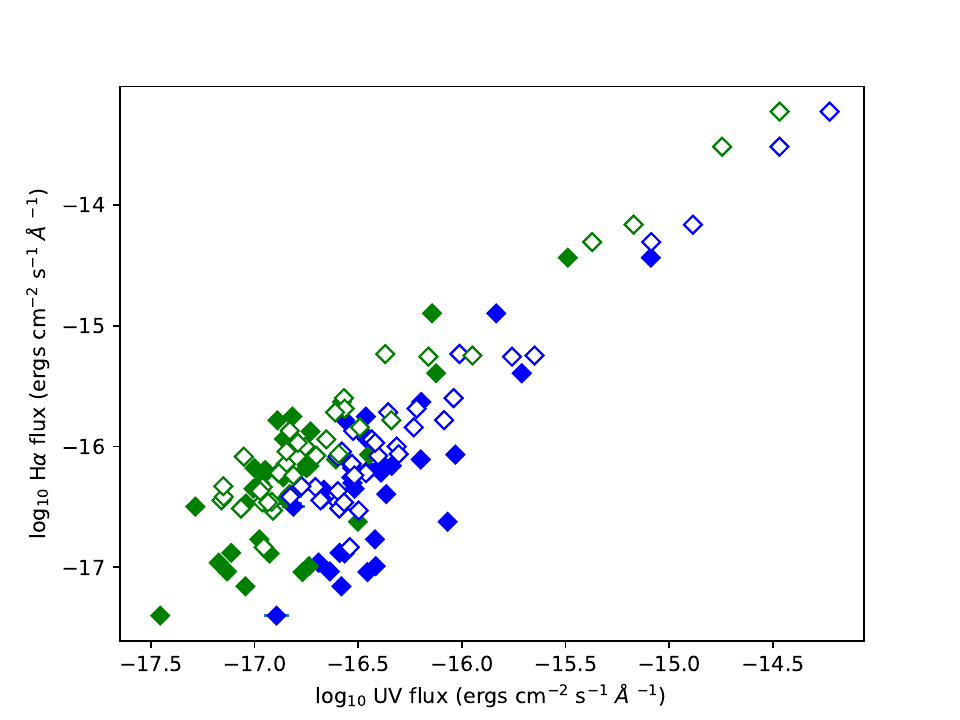}
\caption{The FUV (in blue) and NUV flux (in green) of the 78 knots detected for JO60 is compared to the  $\mathrm{H}{\alpha}$ flux. The 23 knots detected within the disk of the galaxy are shown by the open diamonds, while the filled diamonds are tail knots.
}\label{figure:JO60fuvnuvhalphaknot}
\end{figure}

\begin{figure}
\centering
\includegraphics[width=0.52\textwidth]{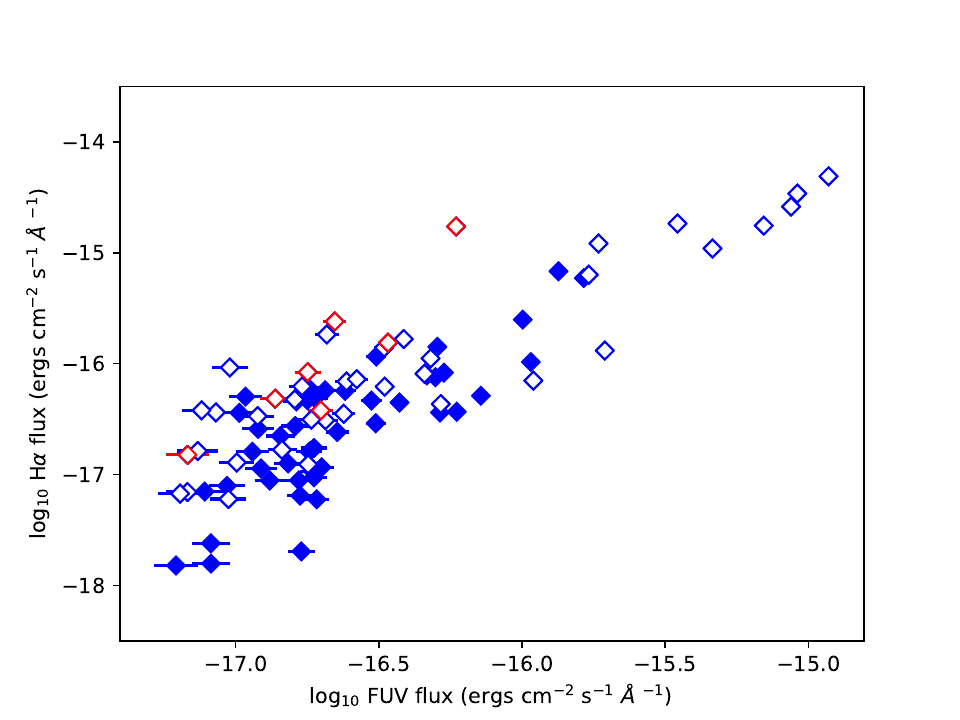}
\caption{The FUV flux of the 87 knots detected for JW39 is compared to the  $\mathrm{H}{\alpha}$ flux. The 42 knots detected within the disk of the galaxy are shown by the open diamonds, while the filled diamonds are tail knots.
The red points are segments from the disk that cover regions with LINER or composite emission as discussed in the text.}\label{figure:JW39fuvhalphaknot}
\end{figure}

\begin{figure}
\centering
\includegraphics[width=0.52\textwidth]{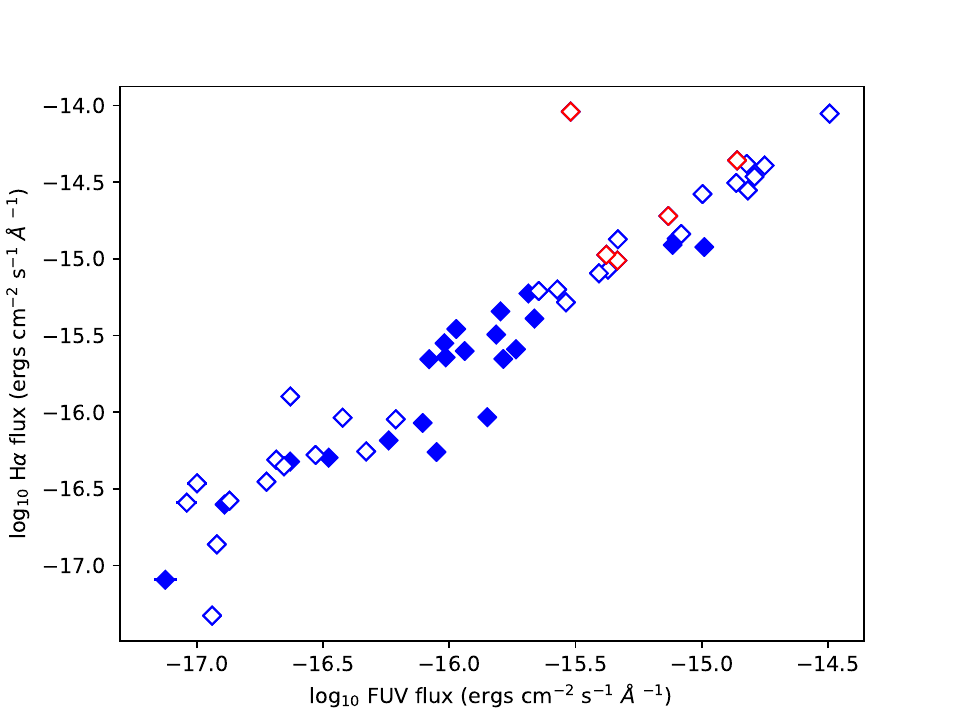}
\caption{The FUV flux of the 54 knots detected for JO194 is compared to the  $\mathrm{H}{\alpha}$ flux. The 32 knots detected from the disk of the galaxy are shown as open diamonds, while the filled diamonds are tail knots. The red points are segments from the disk that cover regions with LINER or composite emission as discussed in the text.}\label{figure:JO194fuvhalphaknot}
\end{figure}

We now compare the UV and H$\alpha$ flux from segments in the three galaxies.  The NUV, FUV flux values are plotted against the H$\alpha$ flux value of JO60 galaxy segments in Fig \ref{figure:JO60fuvnuvhalphaknot} with the blue colour assigned to FUV, green to NUV. The segments on the disk are marked with open diamonds and the segments on the tails are marked with filled diamonds. The FUV flux values are plotted against the H$\alpha$ flux value in Fig \ref{figure:JW39fuvhalphaknot} for JW39 and in Fig \ref{figure:JO194fuvhalphaknot} for JO194. There is a good correspondence between the UV and H$\alpha$ flux values for these segments on the disk and the tail of all three galaxies. The segments on the disk are having a larger area  in general (see Fig \ref{figure:segmap}a, Fig \ref{figure:segmap}b and  Fig \ref{figure:segmap}c) with larger UV and H$\alpha$ flux values as expected.\\

\subsection{Emission line maps \& UV imaging}

Emission line diagnostic ratio diagrams \citep{Baldwin_1981} (BPT) created from MUSE data can be used to distinguish different gas ionization process operating in different regions of the galaxy. We now check whether the regions with UV flux on the disk and tails of the galaxy are contained in star-forming regions in the emission line maps. The BPT maps of JO194, JW39 and JO60 are discussed in detail in \citet{Poggianti_2017b,Peluso_2022,Poggianti_2019b,Radovich_2019} which we used here for our analysis. The $\mathrm{H}{\alpha}$, [SII], [OIII] and [NII] emission line flux maps are used to create the line diagnostic diagrams and distinguish regions dominated by star formation, composite (AGN+SF),  LINER and AGN. The LINER  emission region can in-fact be LINER or LIER. LIER corresponds to non-SF emission coming from the ionisation due to older stellar population or shocks, and spatially corresponds well with diffuse ionised gas \citep{Tomicic_2021a,Tomicic_2021b}. LINER corresponds to emission due to the central black hole feedback.  We show the regions corresponding to LINER (green), AGN (red), Composite (orange) and star formation (blue) along with the NUV or FUV image of the galaxies. JW39 and JO194 host a low-luminosity AGN at the centre, detected as LINER and X-ray emission.

\begin{figure*}
\centering
\subfloat[]{\includegraphics[width=0.35\textwidth]{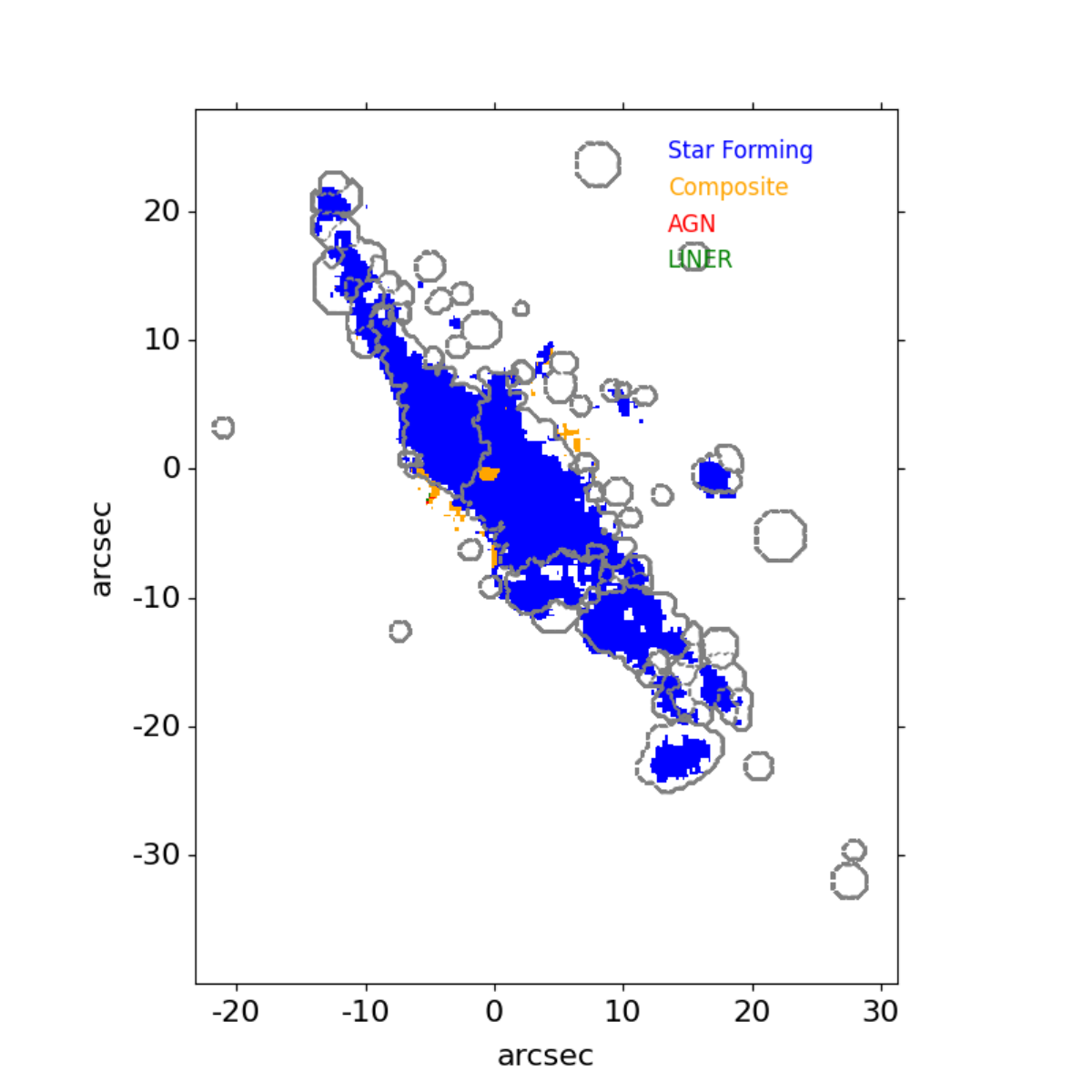}}
\subfloat[]{\includegraphics[width=0.35\textwidth]{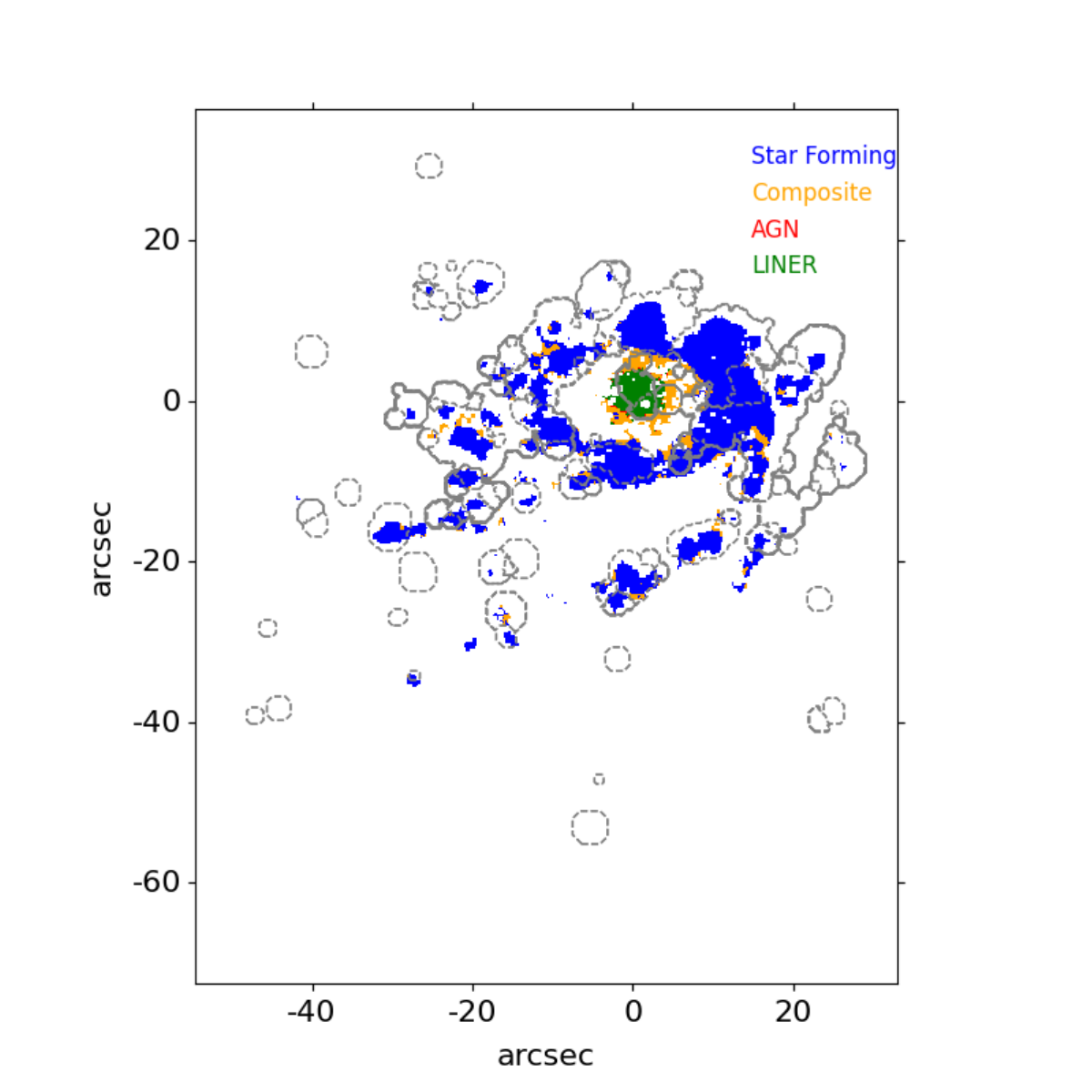}}
\subfloat[]{\includegraphics[width=0.35\textwidth]{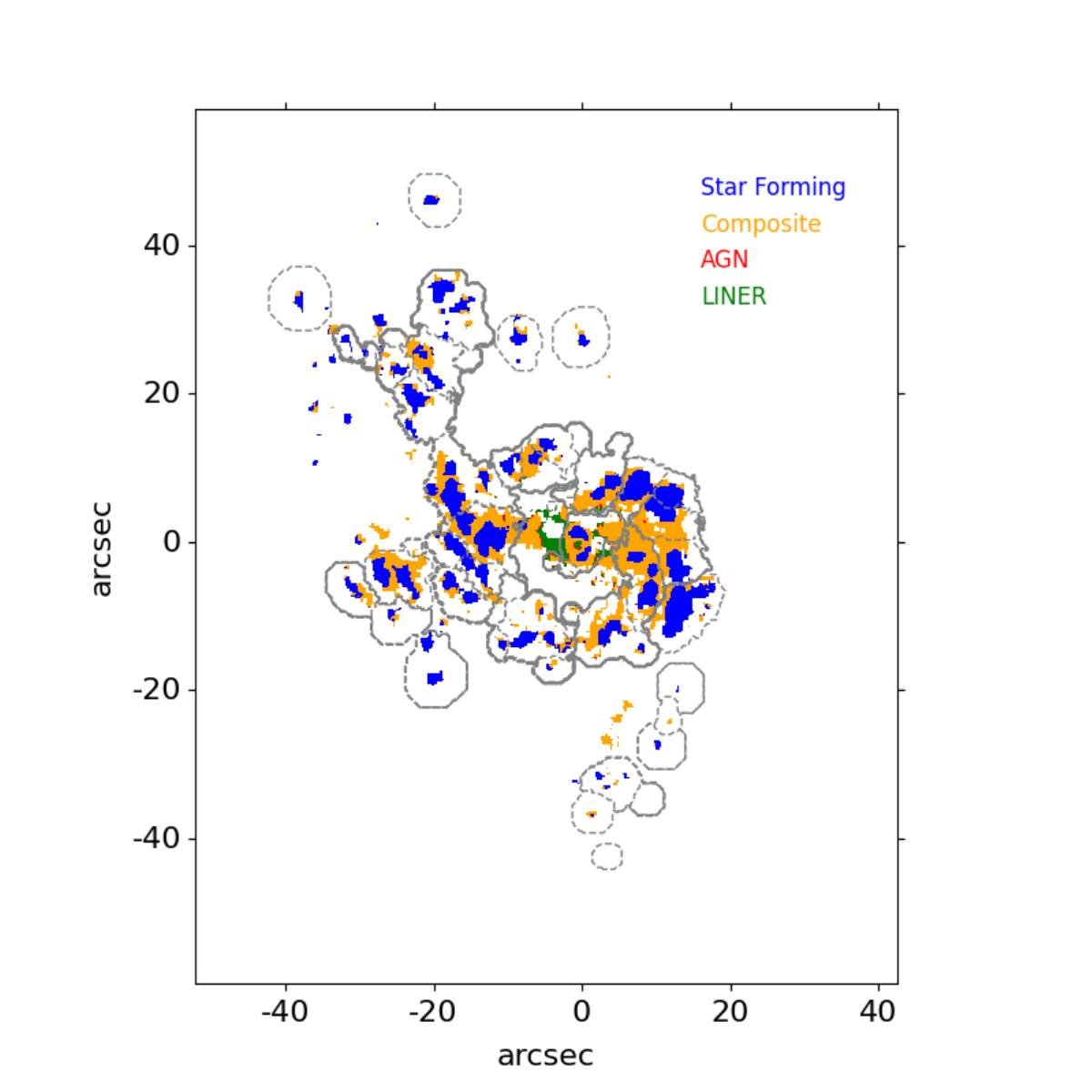}}
\caption{Emission line diagnostic map of galaxy JO60, JW39 and JO194 with regions covered due to emission from LINER, composite (AGN+SF) and star formation. FUV detected segments outline is overlaid in grey.}\label{figure:uvbpt}
\end{figure*}

Figure \ref{figure:uvbpt}a presents the emission line regions for JO60, which is dominated by star formation along the disk and tail of the galaxy. There is a small presence of composite emission at the centre and edges of the disk. The FUV image is showing the presence of UV flux coinciding with the star formation region in the emission line map. We note that the FUV, NUV and H$\alpha$ fluxes from the detected segments of this almost edge-on galaxy are having a good correlation as shown in Fig \ref{figure:JO60fuvnuvhalphaknot}. We also analysed the NUV image of the galaxy and confirm that NUV and FUV shows identical features. Combining the UV, H$\alpha$ and emission line region map thus gives a picture of ongoing star formation happening in the tail and disk of JO60.\\


Figure \ref{figure:uvbpt}b presents the emission line regions for JW39, which is dominated by star formation along the rim of the disk and in the tail of the galaxy. There is  composite and LINER emission in the central region of the galaxy. Interestingly, in this region, there is no significant UV flux other than the point source at the centre, which could be the UV emission from the LINER region. The integrated FUV flux density (Flux/Area) drops by a factor $\sim$ 3 with respect to rest of galaxy disk in an elliptical central region that has a major axis $\sim$ 23 kpc and minor axis $\sim$ 17 kpc. Along the tail there are a few spaxels with composite emission. The FUV image is showing the presence of UV flux coinciding with the star-forming regions in the emission line map. \\


Figure \ref{figure:uvbpt}c presents the emission line regions for JO194. The emission line map shows regions on the disk and the tail with composite and star formation. There is a region close to the centre dominated by LINER emission. The FUV image is showing the presence of UV flux coinciding with the star formation regions in the emission line map. This is consistent both on the disk and the tails in the sense there is a good correspondence between UV flux and star forming region, which is not the case with the composite region. There is not much UV flux at the central region of the galaxy, and in fact there is a "suppressed UV emission" seen which coincides with the composite+LINER region. There is a point source at the centre seen in FUV image and this coincides with the LINER region in the emission line map. The integrated FUV flux density drops by a factor $\sim$ 4 with respect to rest of galaxy disk in an elliptical central disk region that has a major axis $\sim$ 19 kpc and minor axis $\sim$ 14 kpc.

The emission line maps of JW39 and JO194 reveal regions with emission due to composite and LINER. This is seen mostly at the centre for JW39, while for JO194, an almost face-on galaxy, the LINER emission is central and the composite emission is found both at the centre and extending partly along the stripped tails. We checked whether segments with significant contribution from composite and LINER emission deviate from the general UV-$\mathrm{H}{\alpha}$ correlation for star-forming regions. The red points in Fig  \ref{figure:JW39fuvhalphaknot} and Fig  \ref{figure:JO194fuvhalphaknot} correspond to such segments (7 for JW39 and 5 for JO194) from the disk of the galaxy. For both galaxies the segment containing the LINER emission is the offset point with a high H$\alpha$ flux with respect to FUV. The segments with significant contribution from composite region are generally  showing a similar trend comparing to segments detected from star-forming regions in the emission line diagnostic region maps. The segments with contribution from composite/LINER regions are having slightly lower UV flux values for JW39 compared to JO194.

\section{Discussion} \label{sec:Discussion}

 The detected UV flux from the stripped tail and disk of the three galaxies studied here is from the UV integrated spectral energy distribution of hot young OBA spectral type stars on the main sequence. The brightest of these stars will remain on the main sequence for $\sim$ 1-200 Myr which sets an upper limit on the timescale of the 
 star formation visibility as seen in UV. The star formation in the stripped tails of these jellyfish galaxies happens \textit{in situ} and is not formed in the disk of the galaxy prior to stripping (in fact stars in the disk are not affected by ram pressure effects). Thus the UV is directly probing very recent star formation in the tentacles and disk of these jellyfish galaxies. The $\mathrm{H}{\alpha}$ emission is due to the recombination of hydrogen gas ionized by OB spectral type stars with a main sequence life of $<$ 10 Myr \citep{Kennicutt_1998,Kennicutt_2012}. Hence $\mathrm{H}{\alpha}$ and UV are sensitive to different star formation timescales. It is therefore important to understand the features seen in both UV and $\mathrm{H}{\alpha}$ imaging data to make a clear picture of ongoing star formation along the galaxy disk and the tentacles. Stars formed more than 10 Myr ago will not contribute to the $\mathrm{H}{\alpha}$ emission as the O-type stars will have left the main sequence. It is also important to note that the $\mathrm{H}{\alpha}$ emission originates from the ionised gas. There can be mechanisms other than photoionization by  young/massive stars that can ionise the gas, whereas the UV emission is from stellar photospheres. UV is therefore a direct probe to understand star formation in jellyfish galaxies. The good correlation between UV and H$\alpha$ segmentation for the three galaxies presented here demonstrates that there is significant ongoing star formation happening in these galaxies. This star formation can be enhanced in the disk and the stripped tails, following the infall of galaxies into the cluster. The location of most intense UV flux on the disk corresponds to the region of interaction of the galaxy with the ICM. There is a displacement between the galaxy main body traced in optical continuum and the UV emission on the disk where the galaxy likely interacts with the ICM, hinting at the start of quenching process due to ram-pressure stripping on the disk.\\

The scatter in the FUV and H$\alpha$ flux values in Figure \ref{figure:JO60fuvnuvhalphaknot}, Figure \ref{figure:JW39fuvhalphaknot} and Figure \ref{figure:JO194fuvhalphaknot} can in principle be explained by the contribution from other emission mechanisms, but we note that here we didn't perform any extinction correction for both UV and $\rm H\alpha$. We aim to perform extinction correction for a pixel-pixel SFR computation from UV and $\rm H\alpha$, which will be reported in a follow up paper (Tomicic et al. in prep).  The segments on the disk with  contribution from composite emission are not showing any significant offset as described in previous section for JW39 and JO194, while segments that cover the LINER emission show an offset. \\

The UV imaging of the three galaxies presented here shows loose spiral arms which show signatures of arm unwinding, as in the optical. This effect was studied in detail from optical and H$\alpha$ imaging of 11 GASP jellyfish galaxies including JW39 and JO194 by \citet{Bellhouse_2021}. The gas is first stripped from the outer regions of the spiral arm, which then collapses and forms new stars at higher orbits as seen in UV imaging. This increases the pitch angle from the disk and gives the visual appearance of unwinding of spiral arms. We note that the unwinding phenomenon might be rather common in jellyfish galaxies and refer to \citet{Vulcani_2022}, where a catalog of unwinding galaxies is presented and their properties are characterized.\\

The ram-pressure can also drive the gas into the central regions of the galaxy feeding the black hole forming a active galactic nuclei (AGN). This can initiate energetic feedback in the form of outflows, jets into the cold molecular gas needed for sustaining star formation. AGN driven ionized outflows have been observed in GASP jellyfish galaxies \citep{Radovich_2019}. JW39 and JO194 hosts a LINER emitting source in the central region which could be the effect of a black hole fed by gas inflow towards the centre of the galaxy due to ram-pressure stripping \citep{Poggianti_2017b, Peluso_2022}. The outflows and energetic feedback from AGN can inhibit star formation. This is demonstrated in observations and simulations for galaxies in general \citep{Sanders_1988,Springel_2005,Dimateo_2005,Schawinski_2007,Somerville_2008,Hopkins_2008,Schawinski_2010,Wang_2010, Liu_2015,Cheung_2016,Bing_2019} and for jellyfish galaxies in particular \citep{George_2019a,Ricarte_2020}. The emission line diagnostics based on the BPT diagram show an emission dominated by LINER and composite-like emission in the central region of JO194 and JW39. Interestingly this region matches with that of reduced FUV flux. This might indicate that the region with reduced FUV flux corresponds to suppressed star formation due to the effect of LINER dominated ionisation in the same region with a possible connection to AGN feedback. The composite regions in JO194 are too extended to be all likely connected with the AGN. As the LINER emission is concerned, as discussed in \citet{Radovich_2019}, the BPT ratios in JO194 could be explained either by an AGN or by shocks, though the X-ray luminosities measured by Chandra could indicate that an AGN may be indeed present, but possibly obscured. In the case of JO194 there is also a stellar bar present as in the case for JO201. Stellar bars can redistribute gas and quench star formation along the length of the bar \citep{Khoperskov_2018,George_2019b,George_2020} which is further complicated by the presence of AGN. Ram-pressure stripping is an efficient way of removing gas necessary for star formation in the disk of the galaxy. This process also compresses the gas in the disk and triggers starbursts which use the remaining gas in a fast mode. The AGN feedback mechanism operating in the centre of these galaxies undergoing gas stripping will enhance the conditions necessary for star formation quenching. The ram-pressure stripping itself is considered an efficient mechanism to quench star formation for galaxies in groups and clusters. The evidences of suppressed star formation at the centre therefore provides an interesting scenario of accelerated quenching. This can expedite the migration of the galaxy position from the star forming main sequence to the non-star forming passive population.
The presence of UV suppression in jellyfish galaxies seems ubiquitous in all face-on galaxies (JO201, JW39, JO194) studied. This will be further explored in detail with the ongoing UV imaging of a larger sample of GASP galaxies.\\

The stripped tails of jellyfish galaxies are not always dominated by star formation. There exists observations of jellyfish galaxies with long H$\alpha$ tails with no significant UV flux \citep{Boselli_2016,Boissier_2012,Yagi_2007,Yagi_2010,Yagi_2017,Jachym_2017,Laudari_2022}. This implies a different star formation efficiency in the stripped tails and in those cases the H$\alpha$ emission can be produced by other mechanisms for these galaxies. The three galaxies presented here have significant star formation along the tail with a good correspondence between the UV and H$\alpha$ imaging. \\

The UV study of jellyfish galaxies in the local universe can give insights into scenarios of ongoing and suppressed star formation in galaxies undergoing morphological transformation in different environments. The current sample of such galaxies  is limited. The ongoing surveys like GASP with a multi-wavelength data set on jellyfish galaxies with different inclinations can provide more insights into the different phases of ram-pressure stripping, the associated star formation and quenching. This can give important constrains to future surveys of jellyfish galaxies at different redshift in the era of space telescopes like Euclid, JWST and the ground telescopes like E-ELT.

\section{Summary}\label{sec:Summary}

We have studied the ongoing star formation in the GASP jellyfish galaxies JO60, JW39, JO194 using the ultraviolet imaging observations from UVIT. We note that very few UV observations of jellyfish galaxies exist in the literature \citep{Chung_2009, Smith_2010,Hester_2010,Fumagalli_2011,Boissier_2012,Kenney_2014,Boselli_2018,Mahajan_2022}. These are for galaxies in nearby Coma and Virgo galaxy clusters. It is therefore important to study the UV properties of a statistical sample of such galaxies from different clusters. We are observing spectacular cases of GASP jellyfish galaxies in UV, beginning with JO201 and JW100 that were published in \citet{George_2018,Poggianti_2019a}. Here we present UV imaging of three additional GASP galaxies selected based on spectacular nature of tails as evident from optical B-band and $\mathrm{H}{\alpha}$.  We compared the FUV/NUV imaging data with the $\mathrm{H}{\alpha}$ imaging data of galaxies and the following inferences are made. 

\begin{itemize}

\item The main body and the stripped tails for all three galaxies show strong UV emission with the emitting regions in UV and  $\mathrm{H}{\alpha}$ showing a very good correlation. This suggests that the major source responsible for the emission are star-forming regions that contain also stars younger than ~10 Myr.

\item There could be other components responsible for the $\mathrm{H}{\alpha}$ emission as revealed from the emission line maps. The UV emission shows a good correspondence with star-forming and composite regions in the emission line map but not with LINER regions. 

\item Indeed, there are indications of reduced UV flux in the central regions of JO194 and JW39 galaxy disks, which coincide with composite and LINER regions in the optical emission line maps, while JO60 is too edge-on to inspect its central regions. This suggests a suppression of the star formation in the central regions of the disk of JO194 and JW39. This should be taken as indication of outside-in and inside-out quenching happening in tandem in these galaxies. Larger samples of jellyfish galaxies with deep UV imaging will be able to ascertain how common it is to observe a reduced UV (and SF) in the inner half of galactic disks (George et al. in prep.).
\end{itemize}

\section*{Acknowledgements}
This publication uses the data from the AstroSat mission of the Indian Space Research  Organisation  (ISRO),  archived  at  the  Indian  Space  Science  Data Centre (ISSDC).  UVIT  project  is  a  result  of collaboration  between  IIA,  Bengaluru,  IUCAA,  Pune, TIFR, Mumbai, several centres of ISRO, and CSA. Based on observations collected by the European Organisation for Astronomical Research in the Southern Hemisphere under ESO program 196.B-0578 (MUSE). This project has received funding from the European Research Council (ERC) under the European Union's Horizon 2020 research and innovation program (grant agreement No. 833824, GASP project). We acknowledge financial contribution from the contract ASI-INAF n.2017-14-H.0, from the grant PRIN MIUR 2017 n.20173ML3WW\_001 (PI Cimatti) and from the INAF main-stream funding programme (PI Vulcani). This research made use of Astropy, a community-developed core Python package for Astronomy \citep{Astropy_Collaboration_2018}.

\section*{Data Availability}

The Astrosat UVIT imaging data underlying this article are available in ISSDC Astrobrowse archive (https://astrobrowse.issdc.gov.in/astro\_archive/archive/Home.jsp), and can be accessed with proposal ID: G07\_002, G08\_002, A05\_108. The data can also be shared directly on request to the corresponding author.





\begin{thebibliography}{}


\bibitem[Agrawal(2006)]{Agrawal_2006} Agrawal, P.~C.\ 2006, Advances in Space Research, 38, 2989 


\bibitem[Astropy Collaboration et al.(2018)]{Astropy_Collaboration_2018} Astropy Collaboration, Price-Whelan, A.~M., Sip{\H{o}}cz, B.~M., et al.\ 2018, \aj, 156, 123. doi:10.3847/1538-3881/aabc4f


\bibitem[Baldwin et al.(1981)]{Baldwin_1981} Baldwin, J.~A., Phillips, M.~M., \& Terlevich, R.\ 1981, \pasp, 93, 5 


\bibitem[Bellhouse et al.(2017)]{Bellhouse_2017} Bellhouse, C., Jaff{\'e}, Y.~L., Hau, G.~K.~T., et al.\ 2017, \apj, 844, 49 

\bibitem[Bellhouse et al.(2021)]{Bellhouse_2021} Bellhouse, C., McGee, S.~L., Smith, R., et al.\ 2021, \mnras, 500, 1285. doi:10.1093/mnras/staa3298
v

\bibitem[Bing et al.(2019)]{Bing_2019} Bing, L., Shi, Y., Chen, Y., et al.\ 2019, \mnras, 482, 194.



\bibitem[Boselli et al.(2016)]{Boselli_2016} Boselli, A., Cuillandre, J.~C., Fossati, M., et al.\ 2016, \aap, 587, A68. doi:10.1051/0004-6361/201527795

\bibitem[Boselli et al.(2018)]{Boselli_2018} Boselli, A., Fossati, M., Cuillandre, J.~C., et al.\ 2018, \aap, 615, A114. doi:10.1051/0004-6361/201732410


\bibitem[Boissier et al.(2012)]{Boissier_2012} Boissier, S., Boselli, A., Duc, P.-A., et al.\ 2012, \aap, 545, A142. doi:10.1051/0004-6361/201219957







\bibitem[Calzetti et al.(2000)]{Calzetti_2000} Calzetti, D., Armus, L., Bohlin, R.~C., et al.\ 2000, \apj, 533, 682 



\bibitem[Cardelli et al.(1989)]{Cardelli_1989} Cardelli, J.~A., Clayton, G.~C., \& Mathis, J.~S.\ 1989, \apj, 345, 245 



\bibitem[Cheung et al.(2016)]{Cheung_2016} Cheung, E., Bundy, K., Cappellari, M., et al.\ 2016, \nat, 533, 504 

\bibitem[Chung et al.(2009)]{Chung_2009} Chung, A., van Gorkom, J.~H., Kenney, J.~D.~P., Crowl, H., \& Vollmer, B.\ 2009, \aj, 138, 1741 

\bibitem[Cortese et al.(2007)]{Cortese_2007} Cortese, L., Marcillac, D., Richard, J., et al.\ 2007, \mnras, 376, 157 




\bibitem[Di Matteo et al.(2005)]{Dimateo_2005} Di Matteo, T., Springel, V., \& Hernquist, L.\ 2005, \nat, 433, 604. doi:10.1038/nature03335




\bibitem[Ebeling et al.(2014)]{Ebeling_2014} Ebeling, H., Stephenson, L.~N., \& Edge, A.~C.\ 2014, \apjl, 781, L40 


\bibitem[Fasano et al.(2006)]{Fasano_2006} Fasano, G., Marmo, C., Varela, J., et al.\ 2006, \aap, 445, 805 


\bibitem[Fumagalli et al.(2011)]{Fumagalli_2011} Fumagalli, M., Gavazzi, G., Scaramella, R., \& Franzetti, P.\ 2011, \aap, 528, A46 

\bibitem[Fumagalli et al.(2014)]{Fumagalli_2014} Fumagalli, M., Fossati, M., Hau, G.~K.~T., et al.\ 2014, \mnras, 445, 4335 


\bibitem[George et al.(2018)]{George_2018} George, K., Poggianti, B.~M., Gullieuszik, M., et al.\ 2018, \mnras, 479, 4126

\bibitem[George et al.(2019b)]{George_2019a} George, K., Poggianti, B.~M., Bellhouse, C., et al.\ 2019a, \mnras, 487, 3102

\bibitem[George et al.(2019a)]{George_2019b} George, K., Joseph, P., Mondal, C., et al.\ 2019b, \aap, 621, L4. doi:10.1051/0004-6361/201834500

\bibitem[George et al.(2020)]{George_2020} George, K., Joseph, P., Mondal, C., et al.\ 2020, \aap, 644, A79. doi:10.1051/0004-6361/202038810

\bibitem[Ghosh et al.(2021)]{Ghosh_2021} Ghosh, S.~K., Joseph, P., Kumar, A., et al.\ 2021, Journal of Astrophysics and Astronomy, 42, 20. doi:10.1007/s12036-020-09685-0


\bibitem[Girish et al.(2017)]{Girish_2017} Girish, V., Tandon, S.~N., Sriram, S., Kumar, A., \& Postma, J.\ 2017, Experimental Astronomy, 43, 59 

\bibitem[Gullieuszik et al.(2015)]{Gullieuszik_2015} Gullieuszik, M., Poggianti, B., Fasano, G., et al.\ 2015, \aap, 581, A41 

\bibitem[Gullieuszik et al.(2017)]{Gullieuszik_2017} Gullieuszik, M., Poggianti, B.~M., Moretti, A., et al.\ 2017, \apj, 846, 27 


\bibitem[Gullieuszik et al.(2020)]{Gullieuszik_2020} Gullieuszik, M., Poggianti, B.~M., McGee, S.~L., et al.\ 2020, \apj, 899, 13. doi:10.3847/1538-4357/aba3cb



\bibitem[Gunn \& Gott(1972)]{Gunn_1972} Gunn, J.~E., \& Gott, J.~R., III 1972, \apj, 176, 1 




\bibitem[Hester(2006)]{Hester_2006} Hester, J.~A.\ 2006, \apj, 647, 910

\bibitem[Hester et al.(2010)]{Hester_2010} Hester, J.~A., Seibert, M., Neill, J.~D., et al.\ 2010, \apjl, 716, L14 


\bibitem[Hopkins et al.(2008)]{Hopkins_2008} Hopkins, P.~F., Hernquist, L., Cox, T.~J., \& Kere{\v s}, D.\ 2008, \apjs, 175, 356-389 






\bibitem[J{\'a}chym et al.(2017)]{Jachym_2017} J{\'a}chym, P., Sun, M., Kenney, J.~D.~P., et al.\ 2017, \apj, 839, 114





\bibitem[Jaff{\'e} et al.(2018)]{Jaffe_2018} Jaff{\'e}, Y.~L., Poggianti, B.~M., Moretti, A., et al.\ 2018, \mnras, 476, 4753. doi:10.1093/mnras/sty500




\bibitem[Kenney et al.(2014)]{Kenney_2014} Kenney, J.~D.~P., Geha, M., J{\'a}chym, P., et al.\ 2014, \apj, 780, 119 

\bibitem[Kennicutt(1998)]{Kennicutt_1998} Kennicutt, R.~C., Jr.\ 1998, \araa, 36, 189 

\bibitem[Kennicutt \& Evans(2012)]{Kennicutt_2012} Kennicutt, R.~C., \& Evans, N.~J.\ 2012, \araa, 50, 531 

\bibitem[Khoperskov et al.(2018)]{Khoperskov_2018} Khoperskov, S., Haywood, M., Di Matteo, P., et al.\ 2018, \aap, 609, A60. doi:10.1051/0004-6361/201731211


\bibitem[Kumar et al.(2012)]{Kumar_2012} Kumar, A., Ghosh, S.~K., Hutchings, J., et al.\ 2012, \procspie, 8443, 84431N 

\bibitem[Laudari et al.(2022)]{Laudari_2022} Laudari, S., J{\'a}chym, P., Sun, M., et al.\ 2022, \mnras, 509, 3938. doi:10.1093/mnras/stab3280


\bibitem[Lang et al.(2010)]{Lang_2010} Lang, D., Hogg, D.~W., Mierle, K., Blanton, M., \& Roweis, S.\ 2010, \aj, 139, 1782 





\bibitem[Liu et al.(2015)]{Liu_2015} Liu, G., Arav, N., \& Rupke, D.~S.~N.\ 2015, The Astrophysical Journal Supplement Series, 221, 9.








\bibitem[Mahajan et al.(2022)]{Mahajan_2022} Mahajan, S., Singh, K.~P., Postma, J.~E., et al.\ 2022, \pasa, 39, e048. doi:10.1017/pasa.2022.45


\bibitem[Man \& Belli(2018)]{Man_2018} Man, A. \& Belli, S.\ 2018, Nature Astronomy, 2, 695. doi:10.1038/s41550-018-0558-1

\bibitem[Moretti et al.(2014)]{Moretti_2014} Moretti, A., Poggianti, B.~M., Fasano, G., et al.\ 2014, \aap, 564, A138. doi:10.1051/0004-6361/201323098


\bibitem[Moretti et al.(2017)]{Moretti_2017} Moretti, A., Gullieuszik, M., Poggianti, B., et al.\ 2017, \aap, 599, A81 







\bibitem[Owen et al.(2006)]{Owen_2006} Owen, F.~N., Keel, W.~C., Wang, Q.~D., Ledlow, M.~J., \& Morrison, G.~E.\ 2006, \aj, 131, 1974 

\bibitem[Owers et al.(2012)]{Owers_2012} Owers, M.~S., Couch, W.~J., Nulsen, P.~E.~J., \& Randall, S.~W.\ 2012, \apjl, 750, L23 


\bibitem[Peluso et al.(2022)]{Peluso_2022} Peluso, G., Vulcani, B., Poggianti, B.~M., et al.\ 2022, \apj, 927, 130. doi:10.3847/1538-4357/ac4225



\bibitem[Poggianti et al.(2016)]{Poggianti_2016} Poggianti, B.~M., Fasano, G., Omizzolo, A., et al.\ 2016, \aj, 151, 78 

\bibitem[Poggianti et al.(2017)]{Poggianti_2017a} Poggianti, B.~M., Moretti, A., Gullieuszik, M., et al.\ 2017a, \apj, 844, 48 


\bibitem[Poggianti et al.(2017)]{Poggianti_2017b} Poggianti, B.~M., Jaff{\'e}, Y.~L., Moretti, A., et al.\ 2017b, \nat, 548, 304 



\bibitem[Poggianti et al.(2019)]{Poggianti_2019a} Poggianti, B.~M., Ignesti, A., Gitti, M., et al.\ 2019a, \apj, 887, 155

\bibitem[Poggianti et al.(2019)]{Poggianti_2019b} Poggianti, B.~M., Gullieuszik, M., Tonnesen, S., et al.\ 2019b, \mnras, 482, 4466. doi:10.1093/mnras/sty2999


\bibitem[Postma \& Leahy(2017)]{Postma_2017} Postma, J.~E., \& Leahy, D.\ 2017, \pasp, 129, 115002 

\bibitem[Radovich et al.(2019)]{Radovich_2019} Radovich, M., Poggianti, B., Jaff{\'e}, Y.~L., et al.\ 2019, \mnras, 486, 486


\bibitem[Rawle et al.(2014)]{Rawle_2014} Rawle, T.~D., Altieri, B., Egami, E., et al.\ 2014, \mnras, 442, 196 

\bibitem[Robotham et al.(2018)]{Robotham_2018} Robotham, A.~S.~G., Davies, L.~J.~M., Driver, S.~P., et al.\ 2018, \mnras, 476, 3137. doi:10.1093/mnras/sty440


\bibitem[Ricarte et al.(2020)]{Ricarte_2020} Ricarte, A., Tremmel, M., Natarajan, P., et al.\ 2020, \apjl, 895, L8. doi:10.3847/2041-8213/ab9022





\bibitem[Sanders et al.(1988)]{Sanders_1988} Sanders, D.~B., Soifer, B.~T., Elias, J.~H., Neugebauer, G., \& Matthews, K.\ 1988, \apjl, 328, L35 



\bibitem[Schawinski et al.(2007)]{Schawinski_2007} Schawinski, K., Thomas, D., Sarzi, M., et al.\ 2007, \mnras, 382, 1415 


\bibitem[Schawinski et al.(2010)]{Schawinski_2010} Schawinski, K., Dowlin, N., Thomas, D., Urry, C.~M., \& Edmondson, E.\ 2010, \apjl, 714, L108 





\bibitem[Smith et al.(2010)]{Smith_2010} Smith, R.~J., Lucey, J.~R., Hammer, D., et al.\ 2010, \mnras, 408, 1417 

\bibitem[Somerville et al.(2008)]{Somerville_2008} Somerville, R.~S., Hopkins, P.~F., Cox, T.~J., Robertson, B.~E., \& Hernquist, L.\ 2008, \mnras, 391, 481 


\bibitem[Springel et al.(2005)]{Springel_2005} Springel, V., Di Matteo, T., \& Hernquist, L.\ 2005, \apjl, 620, L79 


\bibitem[Subramaniam et al.(2016)]{Annapurni_2016} Subramaniam, A., Tandon, S.~N., Hutchings, J., et al.\ 2016, \procspie, 9905, 99051F 


\bibitem[Tandon et al.(2017a)]{Tandon_2017a} Tandon, S.~N., Hutchings, J.~B., Ghosh, S.~K., et al.\ 2017a, Journal of Astrophysics and Astronomy, 38, 28. doi:10.1007/s12036-017-9445-x


\bibitem[Tandon et al.(2017b)]{Tandon_2017b} Tandon, S.~N., Subramaniam, A., Girish, V., et al.\ 2017b, \aj, 154, 128. doi:10.3847/1538-3881/aa8451

\bibitem[Tandon et al.(2020)]{Tandon_2020} Tandon, S.~N., Postma, J., Joseph, P., et al.\ 2020, \aj, 159, 158. doi:10.3847/1538-3881/ab72a3

\bibitem[Tomi{\v{c}}i{\'c} et al.(2021a)]{Tomicic_2021a} Tomi{\v{c}}i{\'c}, N., Vulcani, B., Poggianti, B.~M., et al.\ 2021a, \apj, 922, 131. doi:10.3847/1538-4357/ac230e

\bibitem[Tomi{\v{c}}i{\'c} et al.(2021b)]{Tomicic_2021b} Tomi{\v{c}}i{\'c}, N., Vulcani, B., Poggianti, B.~M., et al.\ 2021b, \apj, 907, 22. doi:10.3847/1538-4357/abca93











\bibitem[Vulcani et al.(2018)]{Vulcani_2018} Vulcani, B., Poggianti, B.~M., Gullieuszik, M., et al.\ 2018, \apjl, 866, L25. doi:10.3847/2041-8213/aae68b

\bibitem[Vulcani et al.(2022)]{Vulcani_2022} Vulcani, B., Poggianti, B.~M., Smith, R., et al.\ 2022, \apj, 927, 91. doi:10.3847/1538-4357/ac4809


\bibitem[Wang et al.(2010)]{Wang_2010} Wang, J., Fabbiano, G., Risaliti, G., et al.\ 2010, \apj, 719, L208.

\bibitem[Yagi et al.(2007)]{Yagi_2007} Yagi, M., Komiyama, Y., Yoshida, M., et al.\ 2007, \apj, 660, 1209. doi:10.1086/512359

\bibitem[Yagi et al.(2010)]{Yagi_2010} Yagi, M., Yoshida, M., Komiyama, Y., et al.\ 2010, \aj, 140, 1814. doi:10.1088/0004-6256/140/6/1814

\bibitem[Yagi et al.(2017)]{Yagi_2017} Yagi, M., Yoshida, M., Gavazzi, G., et al.\ 2017, \apj, 839, 65. doi:10.3847/1538-4357/aa68e3






\end{thebibliography}








\bsp	
\label{lastpage}
\end{document}